\documentclass[12pt]{iopart}%Rule for one coulombs default
 \usepackage{xspace}
 \usepackage{graphicx}
 \usepackage{subcaption}

\newcommand{\ptt}{\ensuremath{p_{\rm{T}}}\xspace}
\newcommand{\apt}{\ensuremath{\langle p_{\rm{T}}\rangle}\xspace}
\newcommand{\aNch}{\ensuremath{\langle N_{\rm{ch}}\rangle}\xspace}

\newcommand{\ST}{\ensuremath{S_{\rm{T}}}\xspace}

\newcommand{\gevc}{\ensuremath{{\rm GeV}/c}\xspace}

\newcommand{\gev}{\ensuremath{{\rm GeV}}\xspace}
\newcommand{\tev}{\ensuremath{{\rm TeV}}\xspace}

\newcommand{\nch}{\ensuremath{\rm{N_{ch}}}\xspace}

\newcommand{\pp}{\ensuremath{\rm p\!+\!p}\xspace}
\newcommand{\ppbar}{\ensuremath{\rm p\!+\!\bar{p}}\xspace}
\newcommand{\ppb}{\ensuremath{\rm p\!+\!Pb}\xspace}
\newcommand{\pbpb}{\ensuremath{\rm Pb\!+\!Pb}\xspace}

\newcommand{\cms}{\ensuremath{\rm{CMS}}\xspace}
\newcommand{\alice}{\ensuremath{\rm{ALICE}}\xspace}

\newcommand{\eposlhc}{\ensuremath{\rm{EPOS-LHC}}\xspace}
\newcommand{\lhc}{\ensuremath{\rm{LHC}}\xspace}

\newcommand{\pythia}{\ensuremath{\rm{PYTHIA}}\xspace}
\newcommand{\epos}{\ensuremath{\rm{EPOS}}\xspace}

\begin{document}
\title[]{Scaling of kinematical, global observables, energy and entropy densities in \pp,  \ppb and \pbpb collisions from 0.01 to 13 \tev}

\author{ E. Cuautle$^1$, E. D. Rosas$^1$, M. Rodr\'iguez-Cahuantzi$^2$}
\address{$^1$ Instituto de Ciencias Nucleares, Universidad Nacional Aut\'onoma de M\'exico. Apartado Postal 70-543, Ciudad de M\'exico  04510, M\'exico.}
\address{$^2$ Facultad de Ciencias F\'{\i}sico Matem\'aticas, Benem\'erita Universidad Aut\'onoma de Puebla, Av. San Claudio y 18 Sur, Edif. FM9-217, Ciudad Universitaria 72570, Puebla, M\'exico
\eads{\mailto{edgar.dominguez.rosas@cern.ch}}}

%\vspace{10pt}
%\begin{indented}
%\item[]February 2022
%\end{indented}

\begin{abstract}
The multiplicity and average transverse momentum of the charged and identified particles produced in different kinds of colliding systems are an example of global observables used to characterize events over a wide range of energy. Studying these observables provides insights into the collective phenomena and the geometric scaling properties of the systems created in ultra-relativistic \ppb, \pbpb, and even in \pp collisions. The first part of this work presents a study of these variables using different Monte Carlo event generators. It analyzes their sensitivity to find collective phenomena at 0.01, 0.9, 2.76, 7, and 13 \tev, finding a less satisfactory description as the energy decreases. 
The second part analyzes the average transverse momentum of charged hadrons as a function of the multiplicity for \pp, \ppb, and \pbpb data from the \cms and \alice experiments. Comparing with Monte Carlo event generators, we look for a possible scaling law of average transverse momentum scaled to the overlap transverse collision area. 
Additionally, the experimental data are used to compute thermodynamical quantities such as the energy and entropy densities in the Bjorken approach. The results are compared with predictions from \epos and \pythia Monte Carlo event generators. We observe an excellent agreement for \apt from \pp but not for thermodynamical observables, where a sudden rise in a small \apt range resembles the lattice QCD results for the $\epsilon/T^4$ as a function of the temperature; however, only the experimental data from \pp show a kind of saturation.
\end{abstract}
%\pacs{12.38.-t,25.75.-q, 25.75.Nq}
% Deben ser entre 3 y 7
%\keywords{magnetic moment, solar neutrinos, astrophysics}
\submitto{\jpg}
\maketitle
 
% \ioptwocol
%\maketitle

%%%%%%%%%%%%%%%%%%%%%%%%%%%%%%%%%%%%%%%%%%%%%%%%%%%%%%%%%%%%%%%%%%%%%%%%%%%%%%%%%%%%%%%%
\section{Introduction}\label{intro}
%%%%%%%%%%%%%%%%%%%%%%%%%%%%%%%%%%%%%%%%%%%%%%%%%%%%%%%%%%%%%%%%%%%%%%%%%%%%%%%%%%%%%%%%%
The CERN proton anti-proton collider results have shown that the flattening of the average transverse momentum  (\apt) as a function of the charged multiplicity (\nch), observed in the central rapidity region, may serve as a probe for the equation of state of hot hadronic matter. These results constitute a possible signal for a phase transition in hadronic collisions~\cite{VanHove:1982vk}.\\
In general, measurements in the low-momentum regime provide essential information
to describe strong interactions using the non-perturbative region of  Quantum Chromodynamics (QCD); QCD-inspired models are usually implemented in Monte Carlo event generators. Additionally, these kinds of measurements are used to constrain the free parameters of such  models.\\
Furthermore, \apt studies as a function of collision energy and multiplicity distributions  provide more detailed information on the kinematic processes of colliding systems. Thermodynamical quantities describing the system created during the collisions are estimated using the transverse momentum of hadrons~\cite{Sahu:2020mzo}.

Jet production seems to explain the increase of \apt with multiplicity. Jets are responsible for the first rise since hard processes dominate over soft ones in this multiplicity region~\cite{Wang:1988bw}. A second rise seems to appear due to jet production and fragmentation; this behavior is consistent with results obtained at Tevatron energies. However, this behavior is not observed
at higher energies. Moreover, the energy dependence has been studied by the CDF experiment that reported \apt as a function of multiplicity for minimum bias events at 630 and 1800 \gev, observing  and increase with energy and multiplicity  for central pseudorapidities.  It has also been observed that  \apt for soft events is not energy dependent whereas for hard events it increases faster as the energy of the colliding system increases~\cite{Rimondi:2004se}.

The behavior of \apt  is almost linear at low collision energies~\cite{Rossi:1974if} it grows faster for higher energies, where it can be described using a  second order logarithmic polynomial as a function of the collision energy~\cite{Khachatryan:2010xs, Aamodt:2010my}. For identified charged particles the \apt exhibits also a mass dependence.
This behavior can not be  explained~\cite{CMS:2012xvn} by the default tune of the \pythia event generator which overestimates the transverse momentum of ATLAS results~\cite{Aaboud:2016itf} for values larger than 40 in charged particle multiplicity, in the pseudorapidity range $|\eta| < 2.5$. 
Alternative models such as the Color Glass Condensate (CGC)~\cite{mclerran}, where the \apt is a universal function of the ratio of the multiplicity density and the transverse area of the collision (\ST)~\cite{bzdak} qualitatively reproduce the experimental data. However, this geometrical model seems to describe the \apt at larger multiplicities for \pp and \ppb collisions but fails for lower multiplicities, and neither can describe \apt for \pbpb collisions. A possible reason for this poor description is that collective effects may be present in \pp, \ppb, and \pbpb systems,  since the   CGC is just a geometric model, and it is not expected to reproduce the data. It is worth mentioning that in \pp collisions, it has been suggested that there are flow-like~\cite{Khachatryan:2016txc} effects in high multiplicity events at \lhc energies.
The \alice collaboration at the LHC has measured the \apt~\cite{Abelev:2013bla} of charged hadrons from \pp collisions at energies of $\sqrt{s}$ = 0.9, 2.76, and 7 \tev  in the kinematical range  $0.15 < \ptt < 10.0 $ \gevc and $|\eta| < 0.3$. ALICE selected all the charged prompt particles produced in \pp collisions including all decay products, except those from weak decays of strange hadrons. The results show a correlation of \apt with \nch which is stronger for \pp than for \ppb and \pbpb collisions. The observed strong correlation has been attributed within  \pythia to Color Reconnections (CR)  between hadronization strings~\cite{Skands:2010ak,Christiansen:2015yca}, which are directly related with multiple parton interactions.\\
Furthermore, the observed enhancement of \apt versus multiplicity  at low \ptt  is not reproduced by any model~\cite{Aamodt:2010my}.
These kinematic and global  observables have been associated in simulation studies to  QGP formation in proton-proton collisions~\cite{Mangano:2017plv}. The relationship between multiplicity and multiple parton interactions~\cite{Cuautle:2015kra} have been studied in an attemp to explain experimental results.  An incoherent superposition of such interactions would lead to a constant \apt at high multiplicities.\\
In high multiplicity  \pp events,  a medium governed by strong interactions is produced~\cite{ALICE:2016fzo},  thus, the produced matter can be described in terms of quark and gluon degree of freedom. The equation of state, relating the energy and entropy densities, pressure, and temperature of such matter, is of fundamental importance to understand its composition and the static and dynamical properties of the medium created in \pp, \ppb and \pbpb collisions at different energies.
In this work we address these issues. The paper is organized as follows:
A brief presentation of hadronization models used by \pythia and  \epos event generators is provided in section II.  Results and their comparison with experimental data and the average transverse momentum versus multiplicity scaled to the transverse collision area are presented in section III: Subsection III A  describes  results on multiplicity and average transverse momentum versus multiplicity,  and the multiplicity  scaled to the transverse collision area obtained  from event generators.  Subsection III B is devoted  to investigate the equation of state and a possible phase transition of QGP through the analysis of the energy and entropy densities,  scaled by $\apt^{4}$ and  $\apt^{3}$, respectively, for charged and identified particles. Finally, the discussions and conclusions are drawn in section IV.

%%%%%%%%%%%%%%%%%%%%%%%%%%%%%%%%%%%%%%%%%%%%%%%%%%%%%%%%%%%%%%%%%%%%%
\section{Modeling parameters of the hadroproduction}\label{models}
%%%%%%%%%%%%%%%%%%%%%%%%%%%%%%%%%%%%%%%%%%%%%%%%%%%%%%%%%%%%%%%%%%%%%
Hadronization processes are one of the most widely studied topics. Despite these efforts, currently we only have effective theories to describe these phenomena. Different event generators have been developed to investigate hadronization processes in \pp, \ppb , and ion-ion collision systems. Each event generator approaches hadron production mechanisms in a different manner but most of them are based on string fragmentation models at high energies. Some general aspects are briefly described in the following lines:
\pythia hadronization processes include the CR mechanisms providing an alternative final interaction mechanism, modifying some kinematical variables of the hadron-hadron collisions and lepton collisions. These final state interactions allow a better description of the transverse momentum measured by the  \alice experiment.
Color Reconnection is based on the production of hadrons through the interaction probability between partons of low and high $p_T$.
These interacting partons can come from the beam or from the remnant of the beam. They can be quantified by  the probability distribution of reconnecting partons, $P_{rec}(p_{T})$, defined as:

\begin{equation}
\label{Eq.CR}
P_{rec}(p_{T}) =\frac{(R_{rec}\; p_{T0})^{2}}{(R_{rec}\; p_{T0})^{2} + p_{T}^{2}},
\end{equation}
where the range of CR,  $0 \leq R_{rec} \leq 10$, is a phenomenological parameter and  $p_{T0}$ is an energy dependent parameter used to damp the low $p_{T}$ divergence of the $2 \rightarrow 2$ QCD cross section.
Equation ~(\ref{Eq.CR}) used for the description of the color flow, is based on the  Multiple  Parton Interaction model. However,  there are other newer approaches~\cite{Christiansen:2015yqa}, or even older ones such as those developed to describe $W$ bosons in $e^{-} + e^{-}$ processes~\cite{Schwaller:2015gea}, each model contains additional parameters to handle CR effects.\\
Hadron-hadron collisions  consider a spherically symmetric matter distribution within each hadron, $\rho ({\bf x})d^3x = \rho(r)d^3x$. In collision processes, the overlap of the colliding hadrons is a function of the impact parameter ($b$) and is given by:
\begin{equation}
O(b) \propto  \int \int d^3x dt \rho(x,y,z)\rho(x,y,z-\sqrt{b^2-t^2}),
\end{equation}

\noindent
where  different $\rho$ distributions have been studied previously~\cite{Sjostrand:2017cdm}.
Taking a simple Gaussian for this $\rho$, the convolution becomes trivial, but it does not give a good enough description of the data. A better choice of this matter distribution is a double Gaussian function given by:
\begin{equation}
\rho (r) = (1-\beta) \frac{1}{r_1^3} \exp(\frac{r^2}{r_1^3}) +
\beta \frac{1}{r_2^3} \exp(-\frac{r^2}{r_2^3}),
\end{equation}

\noindent
this equation corresponds to a distribution with a small core region of radius $r_2$ and containing a fraction ($\beta$) of the total hadronic matter, embedded in a larger hadron of radius $r_1$. Two free parameters $\beta$ and $r_2/r_1$ are sufficient to give the necessary flexibility.\\
This way of modeling the overlap matter distribution allows the study of hadron-hadron collisions as a function of  impact parameter as for heavy-ion collisions. Thus kinematical variables like \apt in hadron production can also be studied in terms of the impact parameter~\cite{Sjostrand:2017cdm}.

\epos simulates interactions among ions as binary interactions, each represented by a parton ladder; this ladder considers a longitudinal color field, conveniently treated as a relativistic string. The string decays via the production of quark anti-quark pairs creating string fragments usually identified with hadrons ~\cite{Kalus-prl98-152301}.\\
Our \epos \pp simulation at the energy of $\sqrt(s) = 10 $ \gev used the version 1.99, and for larger than 10 \gev, \epos-LHC was used. From now on, these choices are referred to as \epos. Both versions include a simplified treatment of final state interaction but do not include full hydro as is already possible in most recent \epos 3 not released version. \\
\eposlhc is a minimum bias hadronic generator used for heavy-ion collisions and cosmic ray air shower development. However, the goal of this generator is to describe soft particle production ( \ptt $< \sim $ 5 \gevc) for any system and energy in great detail. Moreover, high-density effects leading to collective behavior in heavy-ion collisions are taken into account~\cite{EPOS-LHC} by the fusion model, which means that there is the probability to form a core and a corona, even in proton-proton collisions. The model considers the hadronization part of the secondary particles, and the flow intensity includes different radial flow types that depend only on the total mass of the high density core produced by the overlap of the string segments due to multiple parton interactions in proton-proton or multiple nucleon interactions. The model also includes different radial flow types in the core created in a small volume due to multiple scattering between partons in a single pair of nucleons, which is the case of  \pp collisions.

%%%%%%%%%%%%%%%%%%%%%%%%%%%%%%%%%%%%%%%%%%%%%%%%%%%%%
\section{Results}
Results presented in this work correspond to \pp, \ppb, and \pbpb collisions at different energies, generated with \pythia, and \epos. The data sample size is different for each colliding system, smaller for \pbpb with respect to the \pp, but large enough to understand the statistical fluctuations and for comparison with available experimental data. The analyzed thermodynamical variables were chosen having in mind  \cms and \alice data.

\subsection{Average transverse momentum and multiplicity}
\begin{figure}[h!]
	\begin{subfigure}{0.33\textwidth}
		\includegraphics[width=\linewidth]{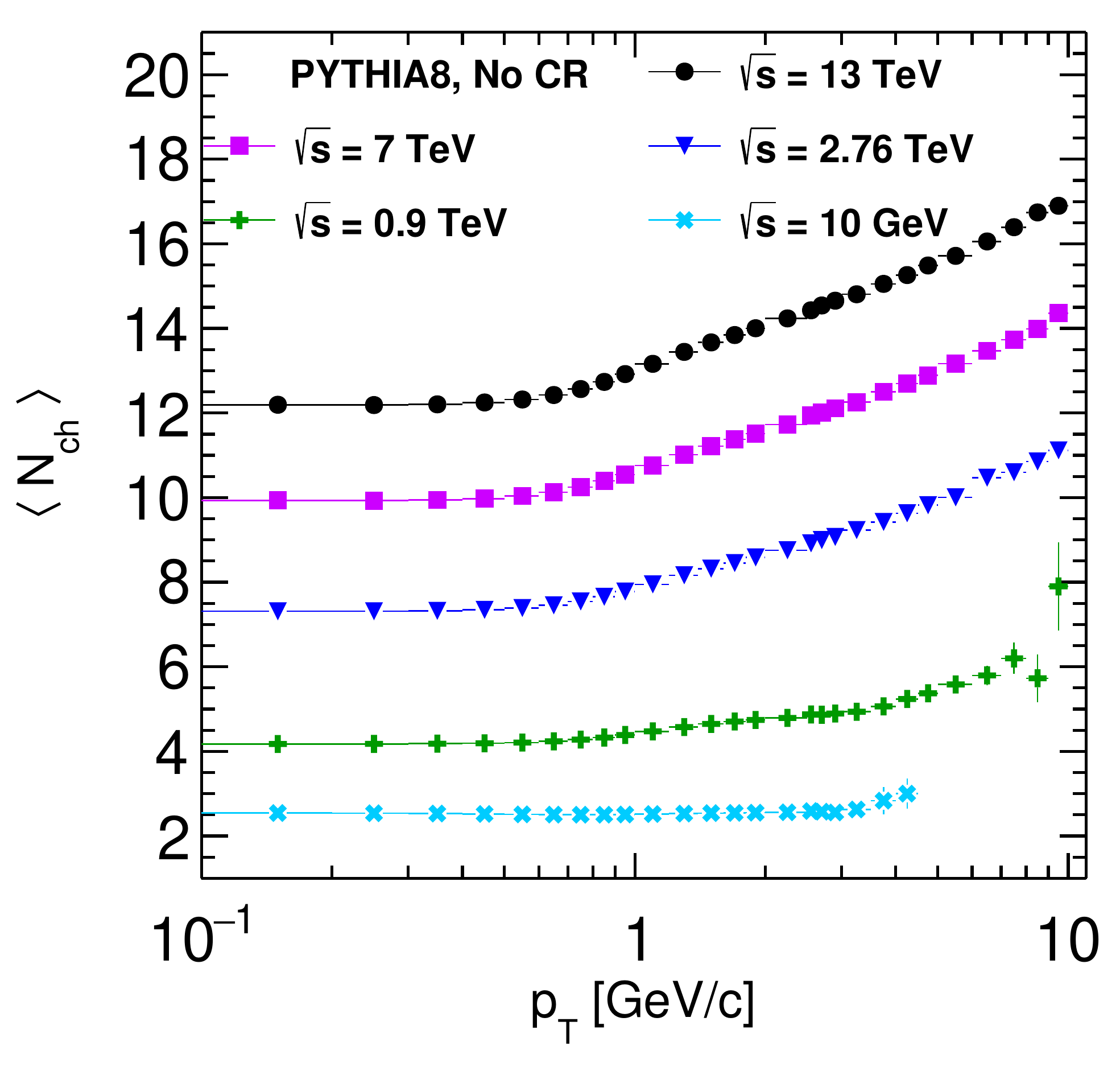}
		\vspace{-0.6cm}
		\caption{} \label{fig:1a}
	\end{subfigure}%
	\hspace*{\fill}   % maximize separation between the subfigures
	\begin{subfigure}{0.33\textwidth}
		\includegraphics[width=\linewidth]{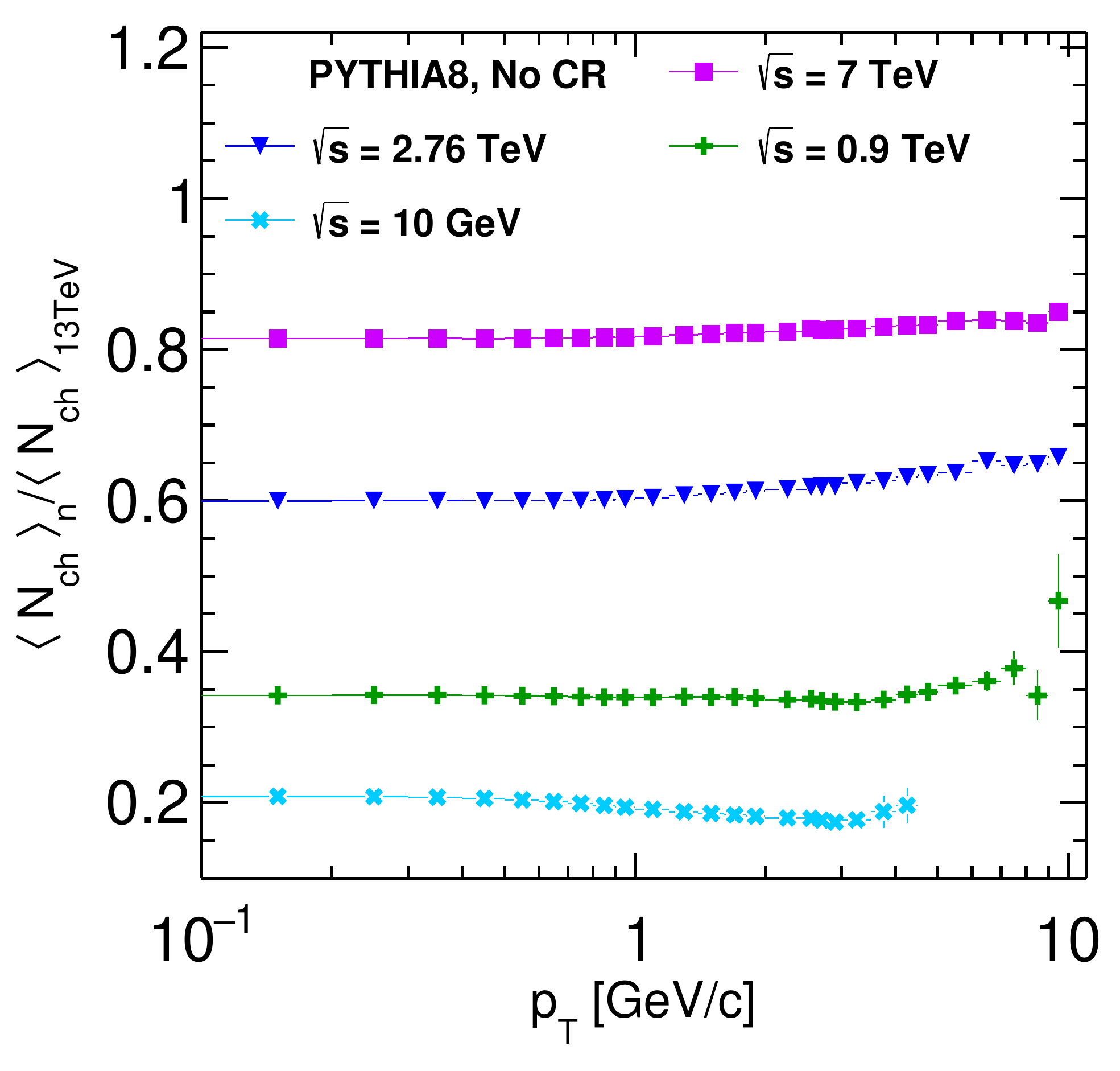}
		\vspace{-0.6cm}
		\caption{} \label{fig:1b}
	\end{subfigure}%
	\hspace*{\fill}   % maximizeseparation between the subfigures
	\begin{subfigure}{0.33\textwidth}
		\includegraphics[width=\linewidth]{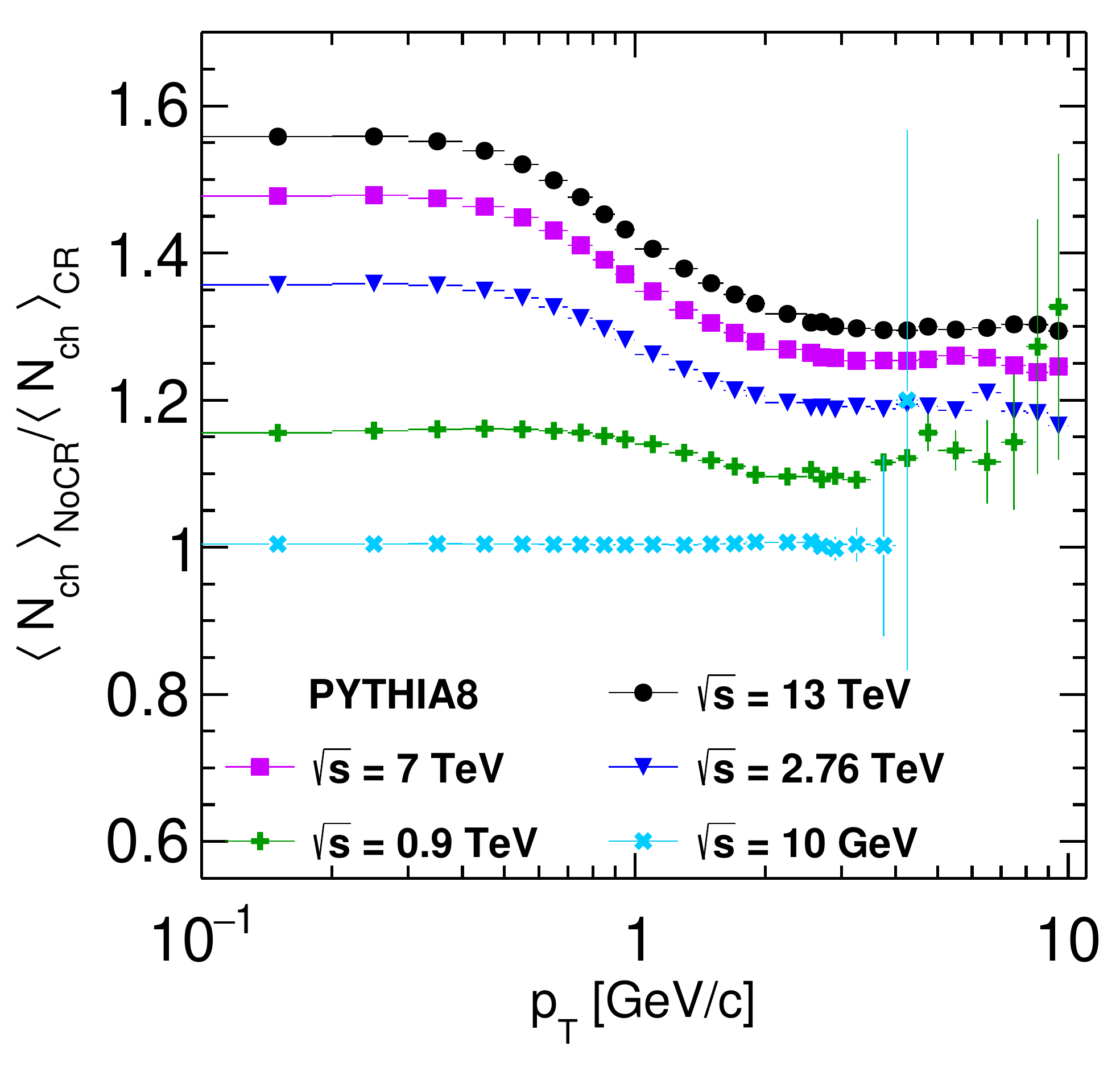}
		\vspace{-0.6cm}
		\caption{} \label{fig:1c}
	\end{subfigure}
	\caption{Average multiplicity distributions as a function of transverse momentum (a) without CR. Ratios of the average  multiplicities  at two energies (b), and ratios between the average multiplicity at the  same energy but without to with  CR (c).} \label{fig:1}
\end{figure}

%{\color{red}
The correlation between multiplicity and transverse momentum has been studied since the 80's~\cite{Ames-Bologna-CERN-Dortmund-Heidelberg-Warsaw:1986amw}. The experimental results on multiplicity suggested the introduction of the multiple parton interaction (MPI)~\cite{Sjostrand:1987su}. Since then, nMPI was introduced in the event generators as an ingredient to describe hadronization processes, together with  the string fragmentation mechanism within the Lund model~\cite{Andersson:1983ia}.\\
%}
Figure~\ref{fig:1} shows the average multiplicity distributions versus transverse momentum generated with \pythia for \pp collisions without CR at different energies in the range from  0.01 to 13 \tev (Fig.~\ref{fig:1} (a)). The distributions at the lowest energy, 10 \gev, show a flat behavior; meanwhile, as the energy increases, a rising slope appears, which can roughly  be described by $\aNch = a + b\cdot \ptt^{c}$, where $a, b$ and $c$ are free parameters. When the energy increases \aNch does as well, similarly to the charged particle density. The exact shape of the average multiplicity is illustrated by looking at  the ratio between average multiplicity for two energies: $\aNch (E_n)$/\aNch(E=13 \tev), where $E_n =0.01, 0.9, 2.76$ and 7 \tev (Fig.~\ref{fig:1} (b)). Figure 1 (c) shows the ratio of the average multiplicity without the CR effect to that where it is  included. The ratio shows  a more prominent difference at higher energy.
For instance, at 13 \tev, the ratio is reduced down to $\approx$ 20\%  when   \ptt goes from $\sim$ 0.5 to $\sim$ 3.0 \gevc. The distributions above 3 \gevc become flat. The CR effects tend to vanish when the energy decreases, producing a null effect at 10 \gev.  These results are presented as a function of the momentum but they also have a dependence on the intensity of the CR, described in Eq. (\ref{Eq.CR}). The transverse momentum distribution experiences a flow-like effect due to  CR~\cite{Ortiz:2013yxa, Bierlich:2017vhg} in \pythia, and it is observed in \pp collisions at \cms~\cite{Khachatryan:2016txc} resembling the distributions observed in heavy ion collisions, where hydrodynamics describes the data qualitatively.\\

\vspace{-0.1cm}

\begin{figure}[h!]
	\begin{subfigure}{0.33\textwidth}
		\includegraphics[width=\linewidth]{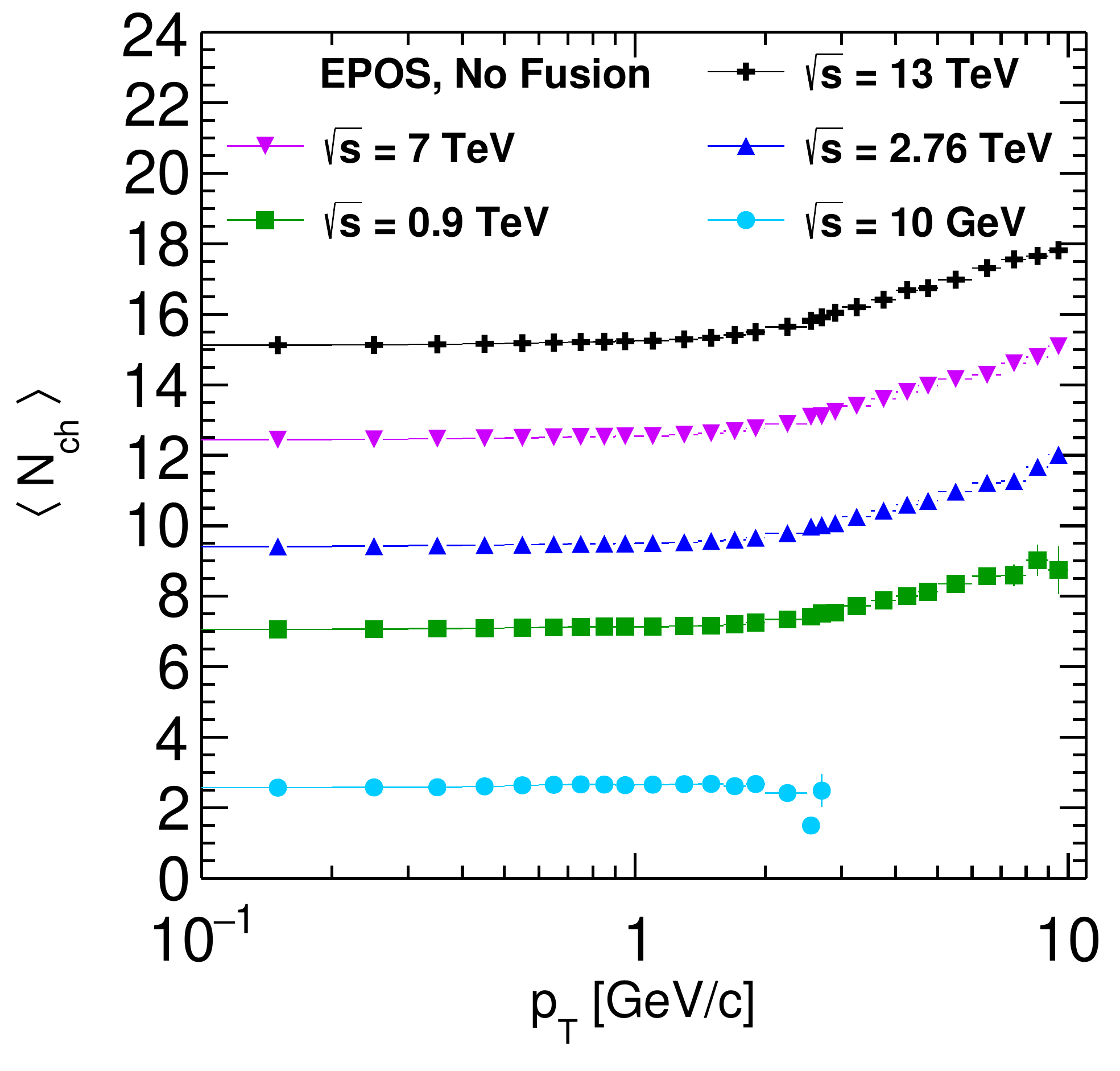}
		\caption{} \label{fig:2a}
	\end{subfigure}%
	\hspace*{\fill}   % maximize separation between the subfigures
	\begin{subfigure}{0.33\textwidth}
		\includegraphics[width=\linewidth]{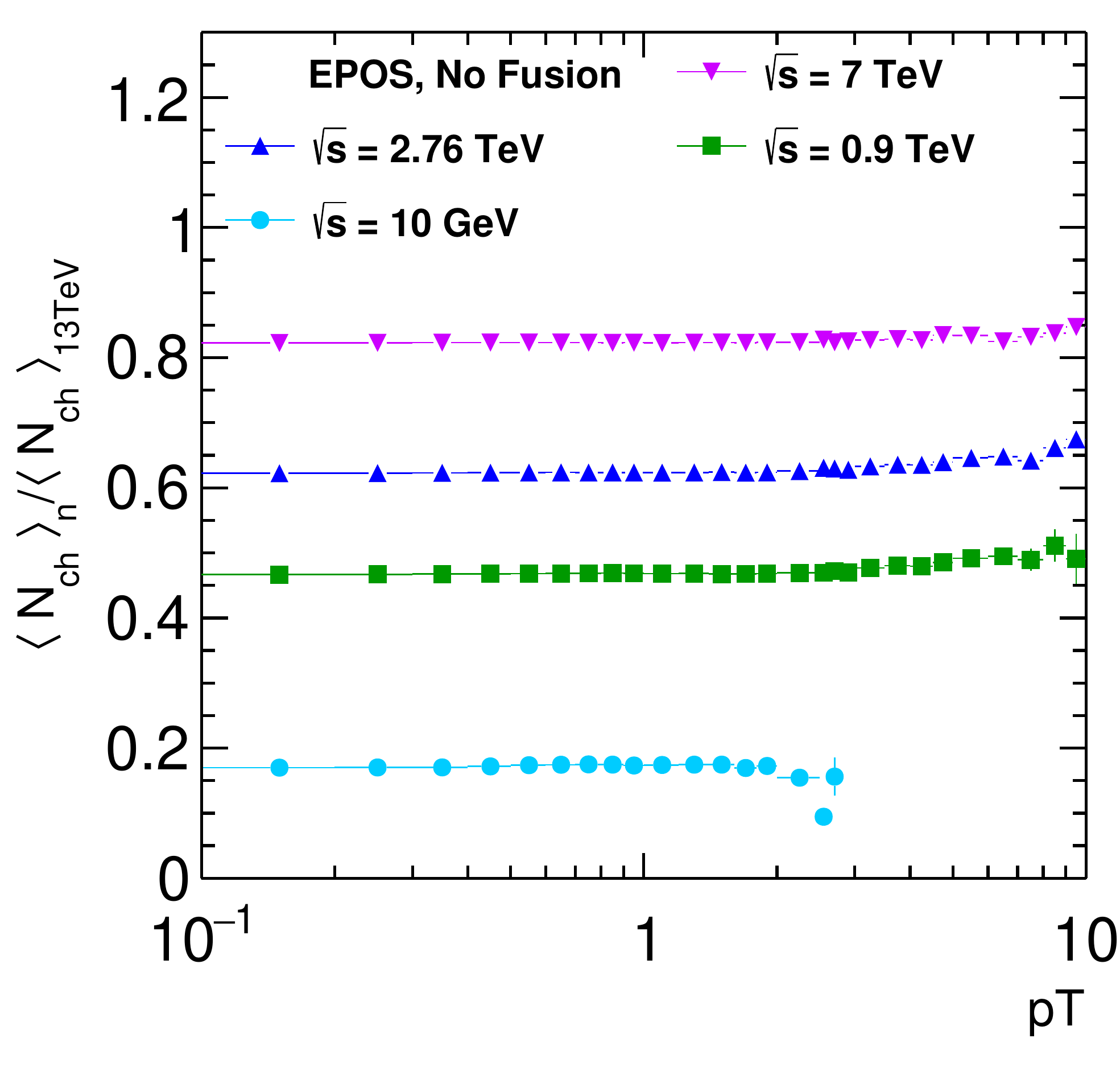}
		\caption{} \label{fig:2b}
	\end{subfigure}%
	\hspace*{\fill}   % maximizeseparation between the subfigures
	\begin{subfigure}{0.33\textwidth}
		\includegraphics[width=\linewidth]{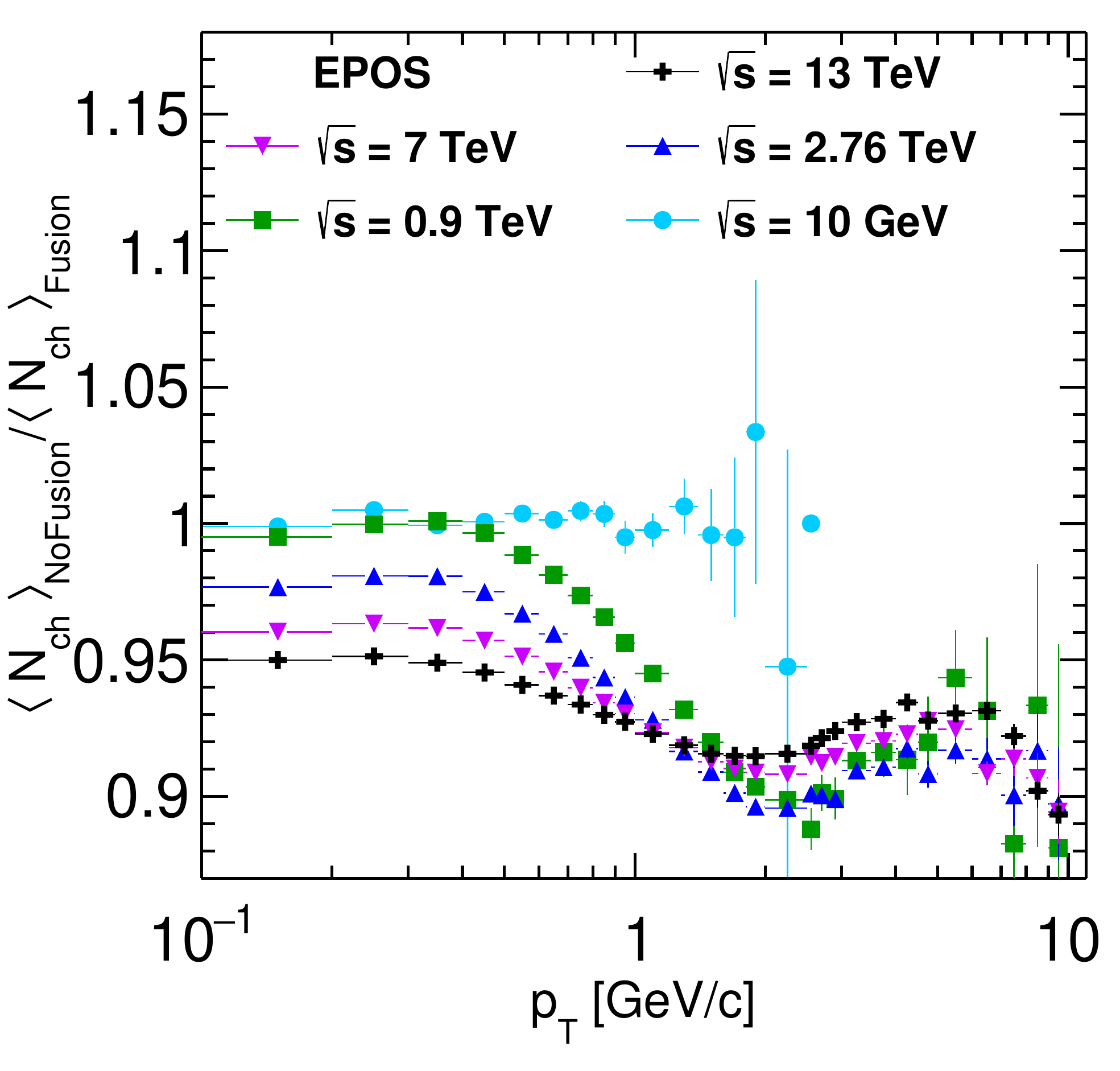}
		\caption{} \label{fig:2c}
	\end{subfigure}
	\caption{
		(color online) Average multiplicity distributions as a function of transverse momentum (a).  Ratios between average multiplicities for two energies (b),   and ratios between average  multiplicities for two models at the  same energy (c).} \label{fig:2}
\end{figure}

%{\color{red}
The distributions in Fig.~\ref{fig:2} are equivalent to the ones shown in Fig.~\ref{fig:1} but computed with the \epos event generator where fusion and no fusion models were considered instead of CR  or no CR included in \pythia. Both Figs. 1 (a) and 2 (a) are very similar, nonetheless in  Fig. 2 (b), \epos produces a flat ratio for 10 \gev whereas \pythia produces ratios that go down and then up.
The ratios of the \aNch given by   fusion off to fusion on models (Fig. 2 (c)) show a slight and constant decrease with increasing energy and  for \ptt larger than $\sim$ 0.5 \gevc,  the ratios decrease except for the case of 10 \gev  which remains constant within uncertainties. Comparing \epos fusion  and \pythia CR, the first predicts a slight reduction on the \aNch whereas the second reduces the \aNch up to  $\sim 20\%$ for higher energy when  \ptt goes from  0.0 - 10 \gevc.

As  mentioned in Sec.~\ref{models}, the impact parameter could modify some kinematic variables like the average transverse momentum versus multiplicity. Figure 3 (a) shows slight differences in the distributions of the impact parameters for \pp collisions incorporated in two event generators, \epos and \pythia. The differences come from the different models to describe the matter distribution of the colliding hadrons and the set of parameters used to fit data. \epos uses  \pythia 6 with  default tune, whereas  \pythia 8 has a different tune and  includes several new functions~\cite{Sjostrand:2004pf} for a  better description of kinematical variables. \\
Our analysis considers the effects on \apt versus multiplicity for two event generators: for the case of  \pythia with   CR, one gets the distributions in  Fig. 3 (b) compared to \alice data~\cite{Abelev:2013bla}, where it is shown that different ranges of impact parameter produce a slight change in the slope of the \apt. Variations in the ranges of the impact parameter by themselves do not allow the reproduction of old data, and this  worsens for low multiplicity,  overestimating by up to $\sim 7\%$ for the impact parameter range of $0.0 < b < 0.2 $ fm, whereas for events with high multiplicity, the statistical errors are large enough to cover the disagreement. Similar variations on the impact parameter are applied to \epos event generator, resulting in a great description of data for the range of $0.0 < b < 0.2 $ fm.  \epos includes hydrodynamical effects; however, kinematic observables from these event generators in hadron-hadron collisions are not the same, nor do they look like the \pythia results discussed in the previous lines.
\begin{figure}[h!]
	
	\begin{subfigure}{0.49\textwidth}
		\includegraphics[width=\linewidth]{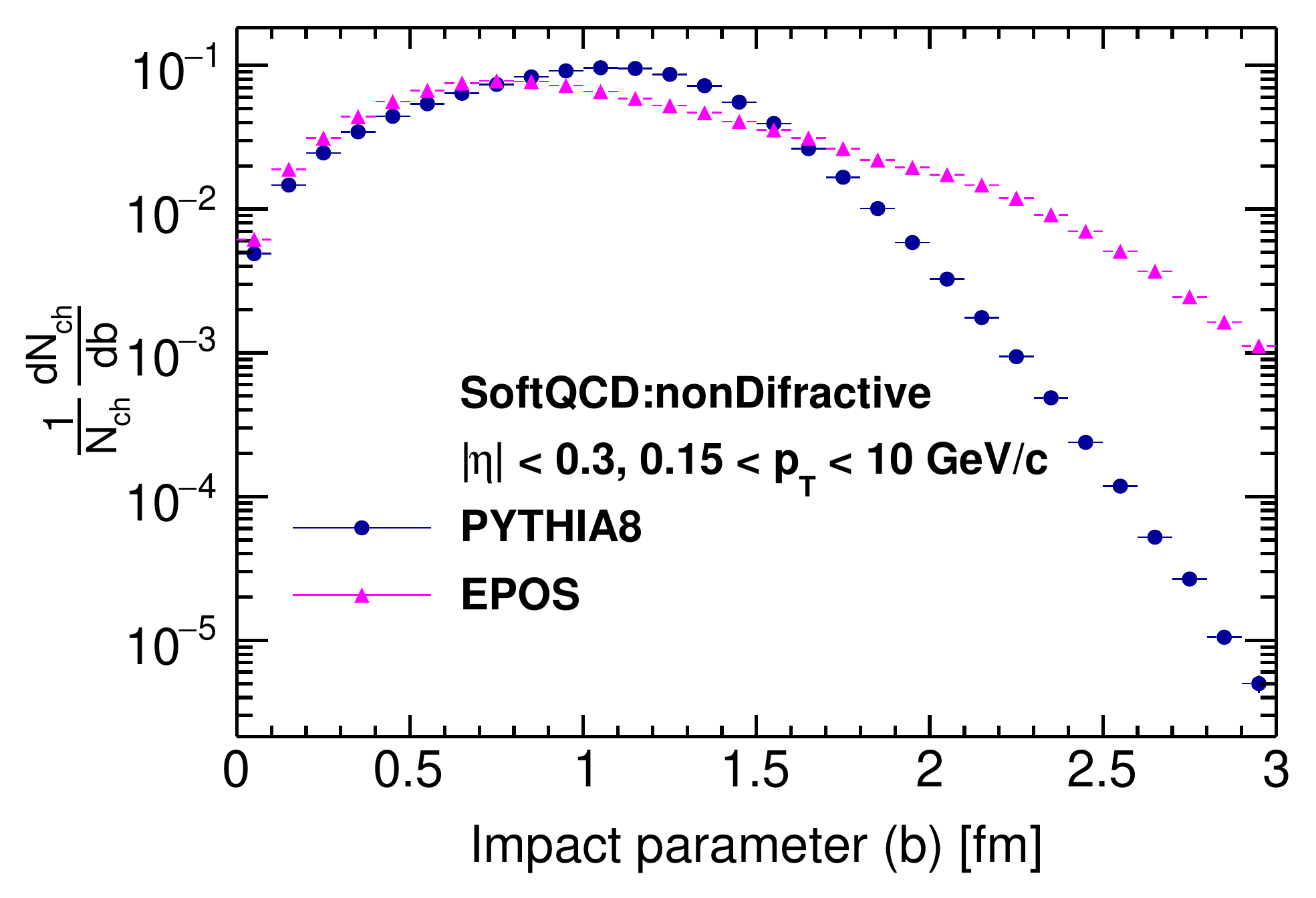}
		\vspace{-0.6cm}
		\caption{} \label{fig:3a}
	\end{subfigure}%
	\hspace*{\fill}   % maximize separation between the subfigures
	\begin{subfigure}{0.49\textwidth}
		\includegraphics[width=\linewidth]{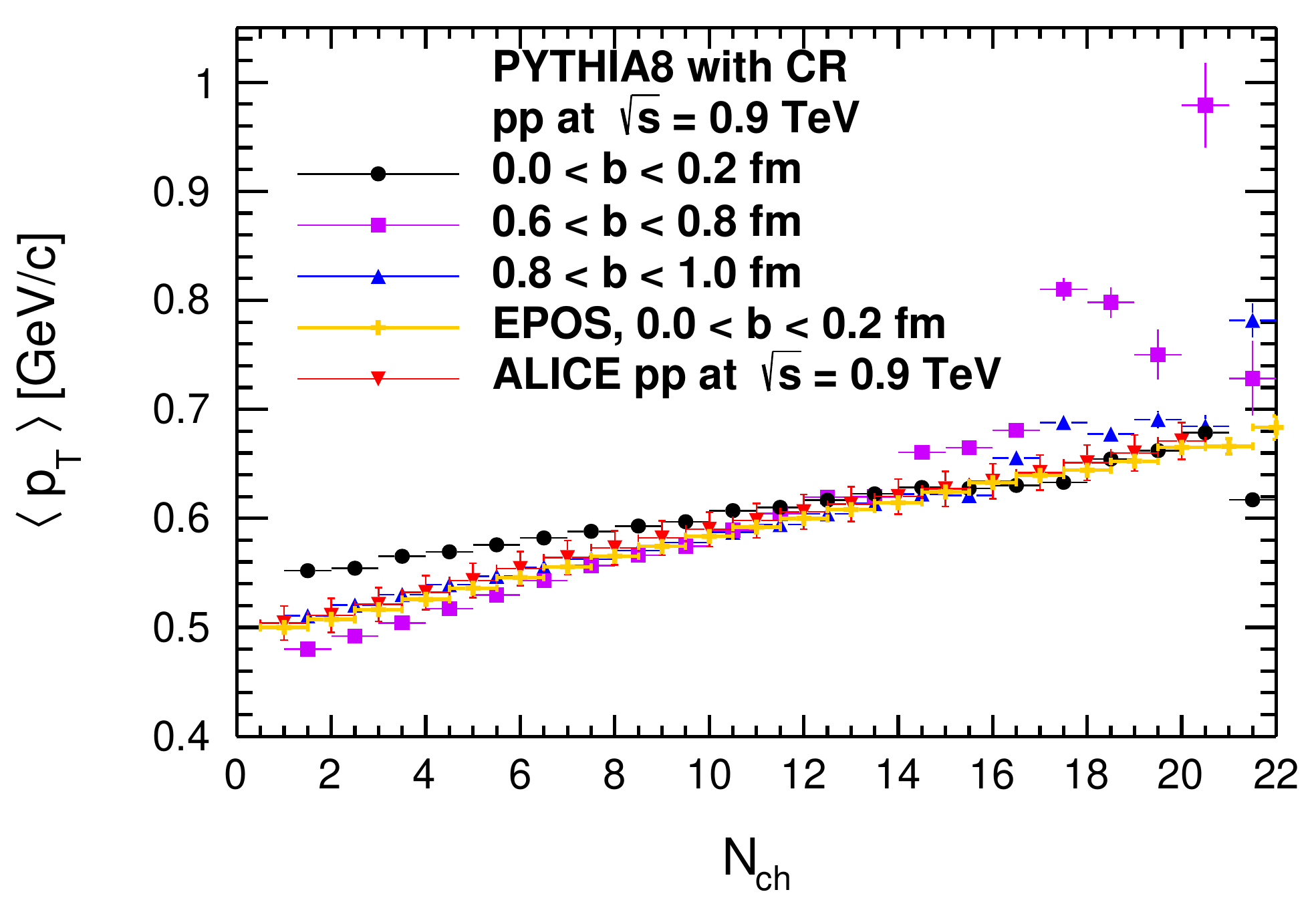}
		\vspace{-0.6cm}
		\caption{} \label{fig:3b}
	\end{subfigure}%
	\caption{impact parameter distributions, from the \epos and \pythia event generators (a), and  average transverse momentum as a function of the multiplicity for three ranges of impact parameter for the \pythia case and one range of \epos,  compared to \alice data ~\cite{Abelev:2013bla} at 900 \gev (b).} \label{fig:3}
\end{figure}\\
This last one includes flow-like effects in hadron-hadron collisions, resembling hydrodynamical effects, but clearly, we can not give a hydrodynamical treatment using this event generator. Figure 4 (a) shows the average transverse momentum versus multiplicity for charged hadrons produced in \pp collisions from two-event generators: \pythia without CR and  \epos with fusion turned off. These results show  a discrepancy for all \ptt ranges. 

\begin{figure}[h!]
	\begin{subfigure}{0.48\textwidth}
		\includegraphics[width=\linewidth]{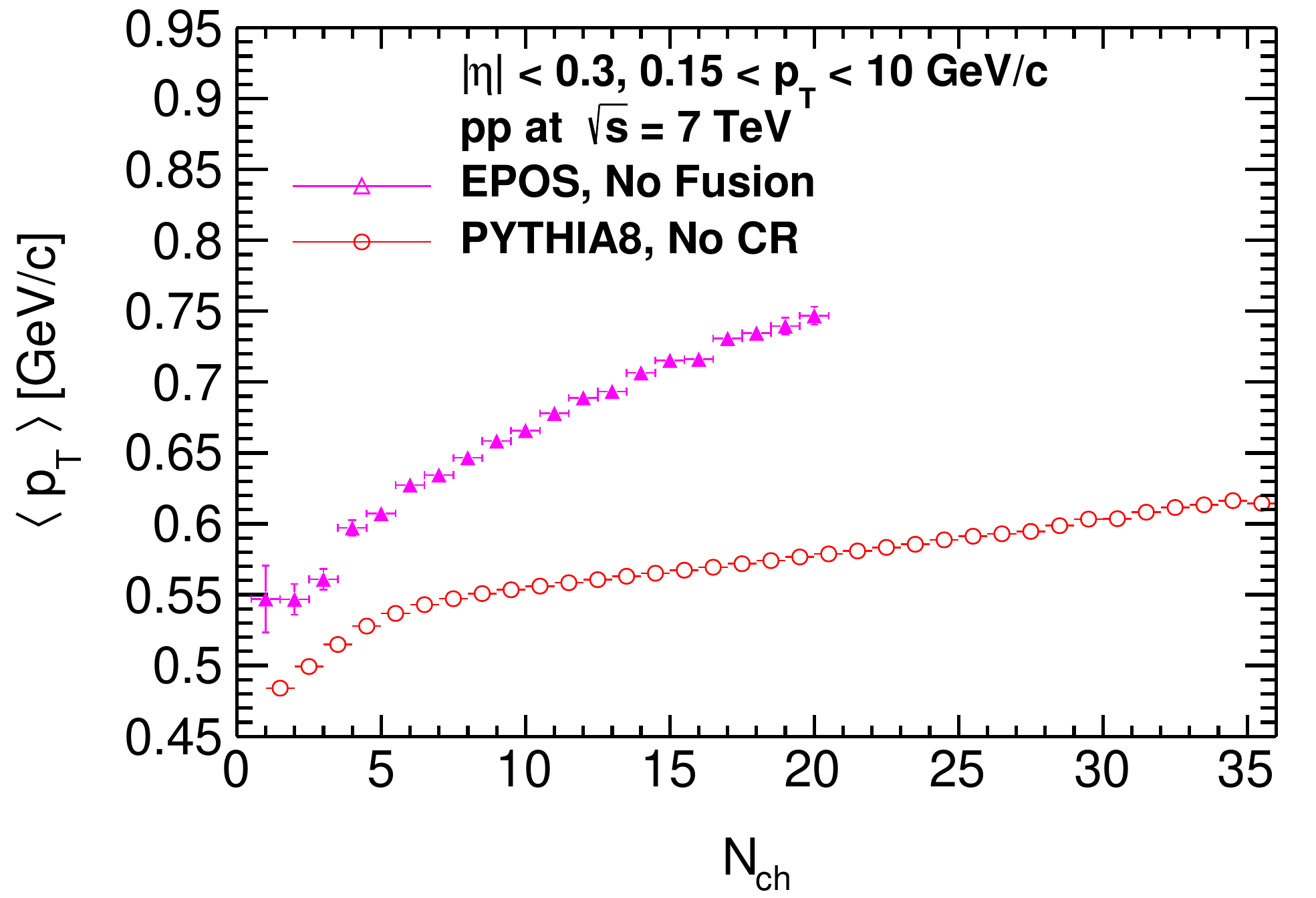}
		\vspace{-0.6cm}
		\caption{} \label{fig:4a}
	\end{subfigure}%
	\hspace*{\fill}   % maximize separation between the subfigures
	\begin{subfigure}{0.48\textwidth}
		\includegraphics[width=\linewidth]{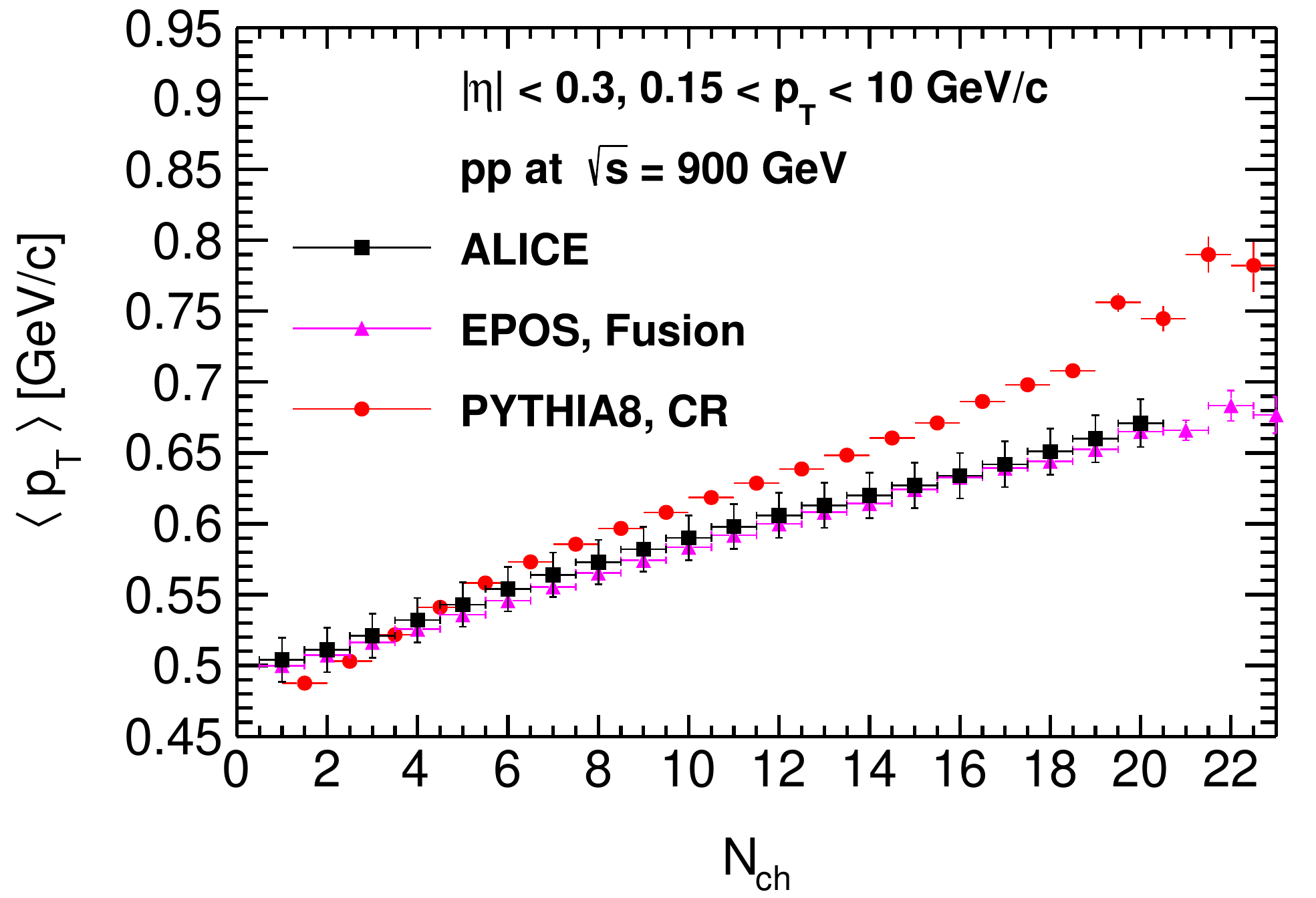}
		\vspace{-0.6cm}
		\caption{} \label{fig:4b}
	\end{subfigure}%
	\vspace{-0.4cm}
	\caption{ Comparison of the average transverse momentum versus multiplicity for two event generators without (a) and with  (b) hydro and  CR of 1.8 is  considered at 900 \gev. \alice data~\cite{Abelev:2013bla} also are showed in the panel (b).}
	\label{fig:4}       % Give a unique label fig.c2
\end{figure}

\begin{figure}[h!]
	\begin{subfigure}{0.47\textwidth}
		\includegraphics[width=\linewidth]{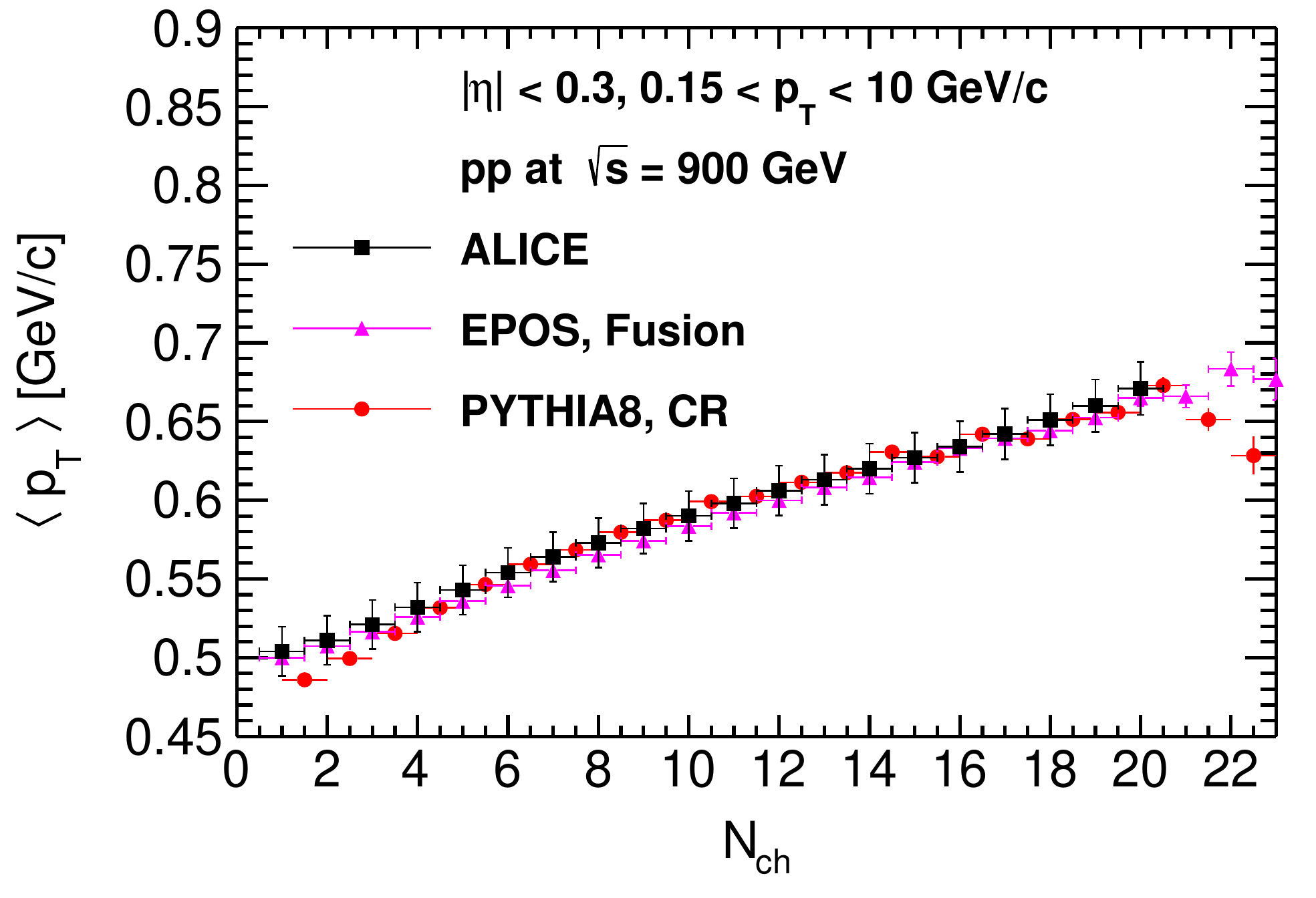}
		\vspace{-0.6cm}
		\caption{} \label{fig:5a}
	\end{subfigure}%
	\hspace*{\fill}   % maximize separation between the subfigures
	\begin{subfigure}{0.47\textwidth}
		\includegraphics[width=\linewidth]{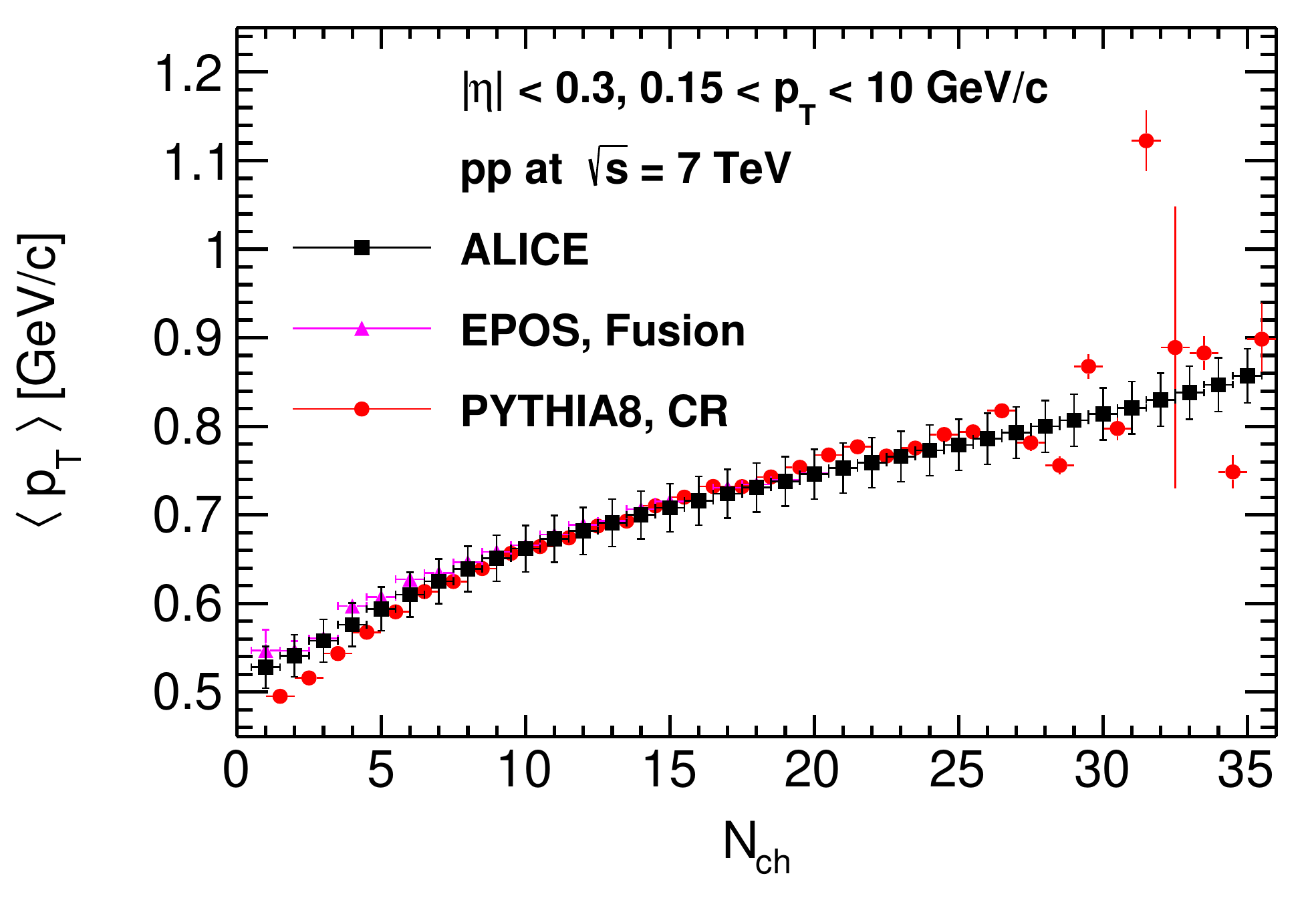}
		\vspace{-0.6cm}
		\caption{} \label{fig:5b}
	\end{subfigure}%
	\vspace{-0.4cm}
	\caption{Comparison of the average transverse momentum versus multiplicity for two event generators  and \alice data~\cite{Abelev:2013bla} at 0.9 \tev (a), and 7 \tev (b).}
	\label{fig:5}       % Give a unique label fig.c3
\end{figure}

\noindent
The same distributions with flow or hydro effects are shown in Fig. 4 (b) together with  \alice data. There is a good agreement between data and \epos for all multiplicity ranges. Nevertheless, \pythia with CR shows an increasing disagreement as the multiplicity increases; besides the model dependency, there is a dependence on energy and the hardness of the process incorporated in each model.
\pythia  and \epos event generators have many parameters which can usually be modified to create a tuning to fit experimental data. Our results of \apt are without tuning but are computed for different impact parameter ranges, which according to section~\ref{models}, could give insight into the hadronic matter distribution in the proton beam.\\
Figure~\ref{fig:5} (a) shows results in agreement among  \pythia, \epos and \alice data at 900 \gev.
Results from \epos considered a range of impact parameter from 0.0-0.3 fm,
since use of the whole impact parameter  range overestimates the \apt by $\sim 9\%$ at low multiplicity.
An  agreement between models and data is also observed for  \apt versus multiplicity at 7 \tev ~\cite{Abelev:2013bla}, showed in Fig.~\ref{fig:5} (b),   where \pythia results were generated with CR parameter of  1.3 instead of 1.8, to  improve the agreement with data up to 6\% for multiplicity below 25,  at least for the \apt from \pp at 7 \tev. At higher multiplicity the agreement worsens.
Results have been tested at different energies but in this work we only show results for 0.9 and 7 \tev. We also tested for \ppb and \pbpb systems with colliding energies  of 5.02 \tev and 2.76 \tev, respectively. These comparisons are complementary to those reported by ALICE~\cite{Abelev:2013bla}.

\noindent
The average transverse momentum for charged hadrons as well as for identified particles has been measured in different colliding systems, since it provides a more detailed understanding of the dynamics of the collisions, like the mass dependence or quark content. The results clearly show that \apt presents the most significant slope for higher mass particles, and gets significantly smaller for the lightest particles~\cite{CMS:2012xvn}. \\
The average transverse momentum as a function of the multiplicity scaled by the transverse area has been proposed as a good scaling law at high multiplicity but not at lower ones. The study of this scaling requires knowledge of interferometry  to extract the size (radius) of the matter created in \pp, \ppb, and heavy-ion collisions.
These techniques were used by \alice~\cite{Abelev:2014pja} and \cms~\cite{Sirunyan:2017ies} to extract the radius as a function of the particle multiplicity. The multiplicity distributions normalized to the transverse area  ($S_{T}$) of the collisions ($N_{ch}/S_{T}$) is an excellent  variable to explore \apt in hadron collisions. The \cms measurements of the multiplicity allow the parametrization of the radius for \pp and \ppb collisions as follows~\cite{McLerran:2013oju}:

\begin{eqnarray}
\label{radii}
R_{pPb, pp} = 1fm \times f_{pPb, pp} \left(^3\sqrt{dN_g/dy}\right)\\
\frac{dN_g}{dy} \approx K \frac{3}{2} \frac{1}{\Delta \eta} N_{tracks},
\end{eqnarray}
where $dN_g/dy$ is the gluon density and

\newcommand\scalemath[2]{\scalebox{#1}{\mbox{\ensuremath{\displaystyle #2}}}}
\begin{equation}
\label{Rpp}
f_{pp}(x) = 
\scalemath{0.9}{
\cases{
	0.387 +0.0335x + 0.274 x^2 -0.0542 x^3  & if  $x < 3.4$ \\
	1.538 & if  $x \geq 3.4$\\}}
\end{equation}

\begin{equation}  %% Will be correc this equation
\label{Rppb}
f_{pPb}(x)   = 
\scalemath{0.9}{
	\cases{
	0.21 +0.47x  & if $ x < 3.5$ \\
	1.184 -0.483x + 0.305x^2 -0.032x^3 & if  $3.5 \leq  x < 5.0$\\
	2.394 & if  $ x \geq 5.0$\\}}
\end{equation}

\noindent
where $N_{tracks}$ is the number of tracks and  $\Delta \eta$ is the pseudorapidity range where the experiment performed its measurements. $K=1$ in our case.
\noindent
%{\color{red}
Figure 6 (a) shows the radii as a function of the multiplicity  computed from Eqs.~(\ref{Rpp}) and~(\ref{Rppb}) for \pp and \ppb collisions, respectively, with the  multiplicity from \alice  \pp,  \ppb, and \pbpb colliding systems \cite{Abelev:2013bla}, and compared to \cms \pp~\cite{CMS:2010qvf} and \ppb~\cite{CMS:2013pdl} results. The parametric distribution measured directly from  \alice~\cite{Abelev:2014pja} has also been presented for the case \pp and \ppb. It corresponds to the lower distribution in the plot.  An agreement at low multiplicity is observed, but the radius increases faster with the multiplicity for the heaviest systems.
As will be discussed later, these differences bring consequences in other calculations, such as energy density and entropy density.
%}\\
\noindent
The average transverse momentum also presents a scaling law when it is plotted as a function of multiplicity scaled by the  transverse area.  Figure 6 (b), shows the comparison among results from \alice  \pp at 7.0 \tev and \pythia with CR  and \epos with fusion turned on,  at the same energy. Results  from \alice \ppb at 5.02 \tev are compared to \epos simulations at the same energy. These scaling laws with experimental data were studied~\cite{McLerran:2013oju} in the framework of CGC, suggesting that no scaling is presented at low \ptt due to flow effects. Nevertheless, our simulated results include flow, showing that the breaking scaling still remains.

\begin{figure}[h!]
	
	\begin{subfigure}{0.49\textwidth}
		\includegraphics[width=\linewidth]{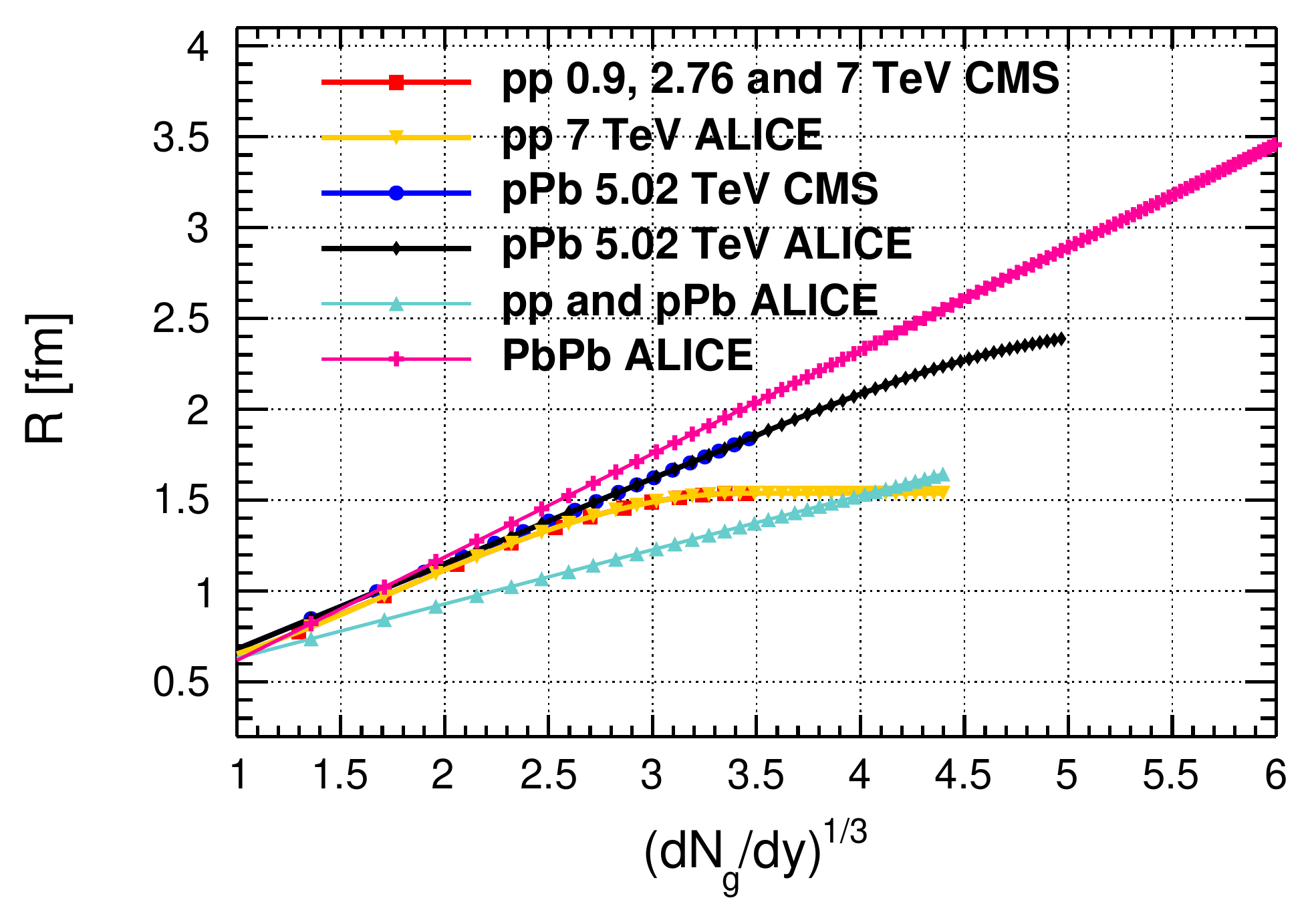}
		\vspace{-0.6cm}
		\caption{} \label{fig:6a}
	\end{subfigure}%
	\hspace*{\fill}   % maximize separation between the subfigures
	\begin{subfigure}{0.49\textwidth}
		\includegraphics[width=\linewidth]{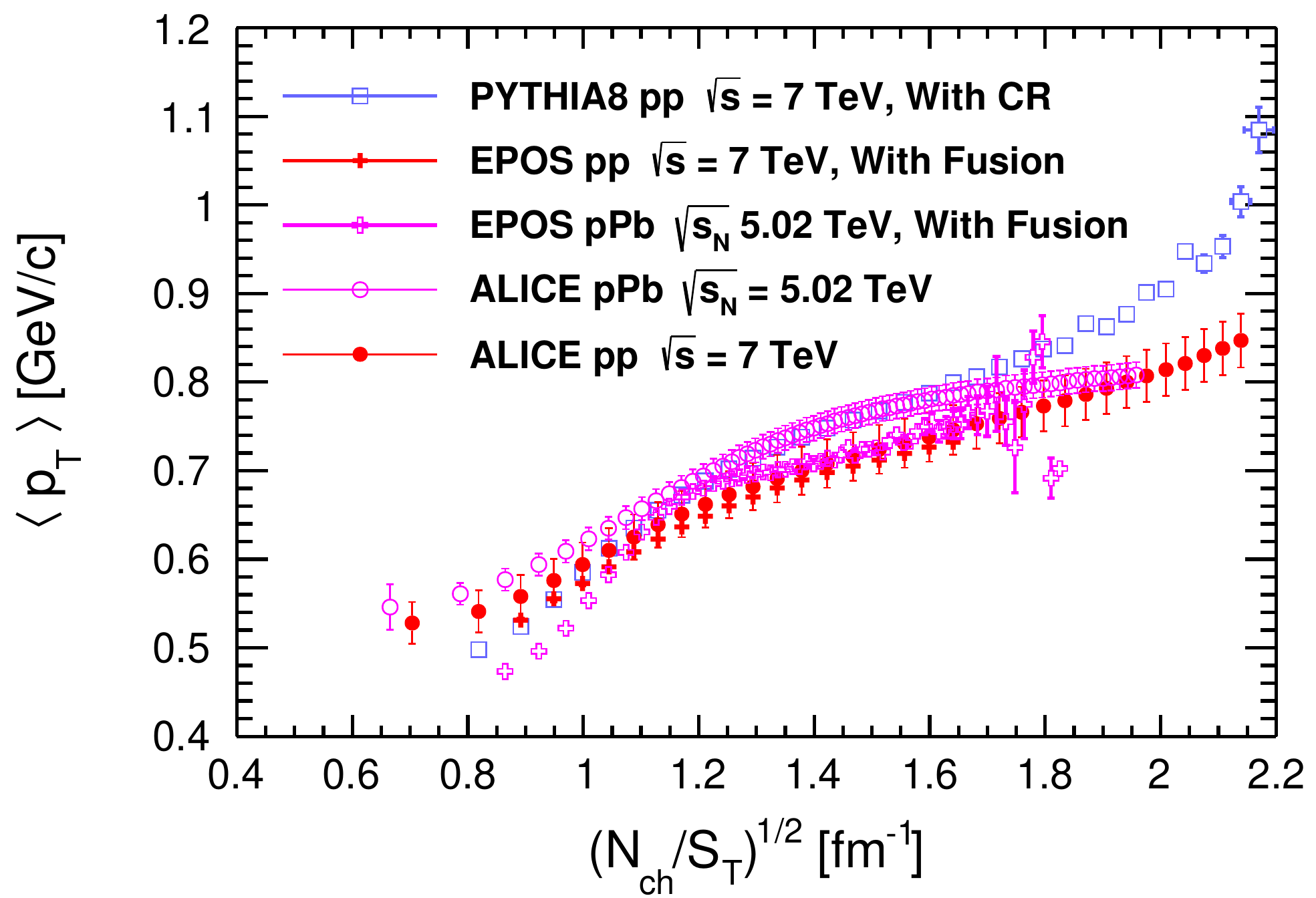}
		\vspace{-0.6cm}
		\caption{} \label{fig:6b}
	\end{subfigure}%

	\caption{Radii as a function of multiplicity according to parametric function~\cite{bzdak},  comparing results from  \alice data~\cite{Abelev:2013bla,Abelev:2014pja},  \cms   \pp~\cite{CMS:2010qvf} and  \ppb~\cite{CMS:2013pdl} data (a).
	Average transverse momentum as a function of multiplicity normalized by transverse area (b), comparing  results from \alice experiment~\cite{Abelev:2013bla} and simulation from \pythia  and \epos event generators.}
	\label{fig:6}       % Give a unique label
\end{figure}

%%%%%%%%%%%%%%%%%%%%%%%%%%%%%%%%%%%%%%%%%%%%%%%%%%%%%%%%5
\subsection{Experimental equation of state in \pp and \pbpb collisions}
\vspace{-0.1cm}
Thermodynamical quantities like energy and entropy densities, temperature, and pressure,  are related by an equation of state to describe the evolution of the medium created in heavy-ion collisions and also in high multiplicity \pp events. Hydrodynamic theory has been used to get the energy density, entropy, temperature and other quantities as a first description approach to \pp collisions~\cite{Chadha:1974qs}. Different observables are analyzed to explore the formation of QGP-droplets in \pp collisions~\cite{Sahoo:2019ifs}.  The HotQCD Collaboration has developed an equation of state in (2+1) flavor QCD~\cite{HotQCD:2014kol} which gives a prediction for the energy density divided by temperature $T^4$ ($\epsilon /T^4$) as a function of $T$. Similar results have been obtained~\cite{Braun:2015eoa} using the Color String Percolation Model, which allow to compute hydrodynamical properties~\cite{Sahu:2020mzo} in  \pp, \ppb and \pbpb collisions.\\
Analyzing data from the \pp and \ppbar collisions, the equation of state can be estimated~\cite{Campanini:2011bj} by  combining  \apt versus multiplicity distribution measurements.\\ Results from heavy-ion collisions can be well described by hydrodynamics, however, this theory does not apply to results from \pp collisions. Nevertheless,   \pp collisions at high  energy produce a high particle density, and the mean free path of particles in such a system is very small compared to its dimension. The created system expands, and a fraction of the expansion processes must have hydrodynamical characteristics. Under this consideration the initial entropy density estimated by Landau~\cite{Landau:1953gs} is proportional to the multiplicity ($s_0 =k dN_{ch}/d\eta$, where $k$ is a constant) which allows to compute the entropy~\cite{Zakharov:2013gya} for a mini-QGP system.
Statistical models can be used to analyze the multiplicity densities and then  get thermodynamical limits, in \pp collisions. In particular, the $\epsilon/T^4$ as a function of multiplicity tends to the same value~\cite{Sharma:2018uma} when it is calculated from different statistical ensembles. The shear viscosity, isothermal compressibility and speed of sound have also been calculated~\cite{Sahu:2020swd} from \alice \pp results.

This work presents the extraction of the thermodynamical properties from  \cms and \alice data, starting from the average transverse momentum as a function of multiplicity. At the same time, the results are compared to those from event generators.

\noindent
The Bjorken energy density ($\epsilon_{Bj}$)~\cite{Bjorken:1982qr} defined by Eq.~(\ref{Eq.Bjorken}) has been used to show that the enhancement of the strangeness production as a function of energy density for different colliding systems presents a scaling law ~\cite{Cuautle:2016huw,Paic:2016jep}. 

\begin{equation}
\label{Eq.Bjorken}
\epsilon_{Bj} \simeq \frac{3}{2} \frac{  \langle p_T \rangle \frac{dN}{d\eta}}{ S_T \cdot \tau}.
\end{equation}

\noindent
In this equation,  \ST is the interaction area calculated as in the previous section, and $\tau$ is the proper time in the Bjorken model, so the product $\ST \cdot \tau$ is the volume occupied by the particles. Using Eq.~(\ref{Eq.Bjorken}), we compute the energy density for charged hadrons from \pp, \ppb, and \pbpb data at different energies measured by \alice. The results as a function of the multiplicity scaled by \ST are shown in Fig. 7 (a), where a clear scaling law is observed at low multiplicity for all colliding systems. At high multiplicity, the energy density shows differences that may come from the measured radii. Analysis of the energy density for identified particles produced in \ppb is shown in  Fig. 7 (b), where higher values for the heaviest particles are evident. It is possible to add the energy by  particle species and the total approximately corresponds to the results for all charged hadrons.

\begin{figure}[h!]
	\begin{subfigure}{0.49\textwidth}
		\includegraphics[width=\linewidth]{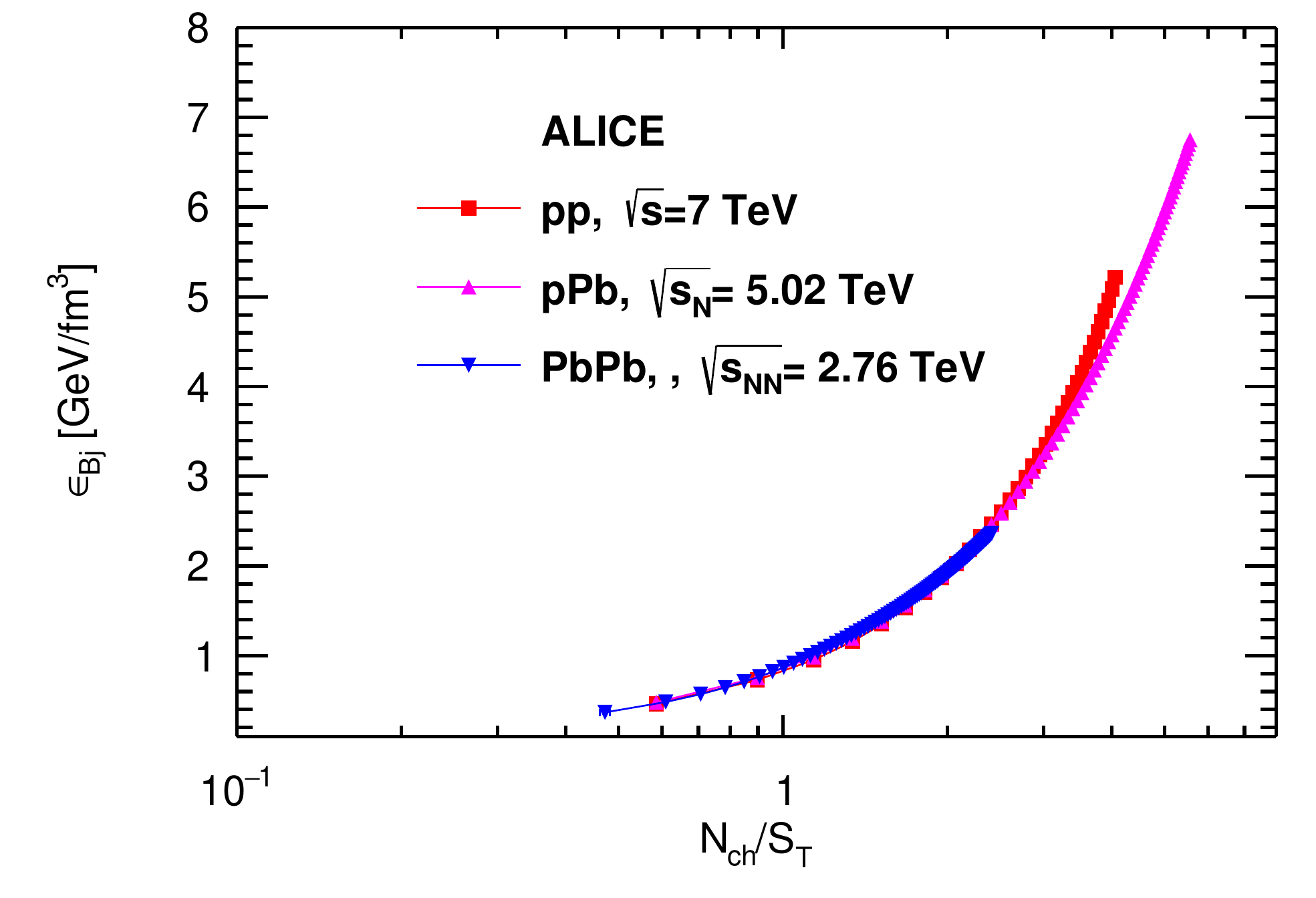}
		\caption{} \label{fig:7a}
	\end{subfigure}%
	\hspace*{\fill}   % maximize separation between the subfigures
	\begin{subfigure}{0.49\textwidth}
		\includegraphics[width=\linewidth]{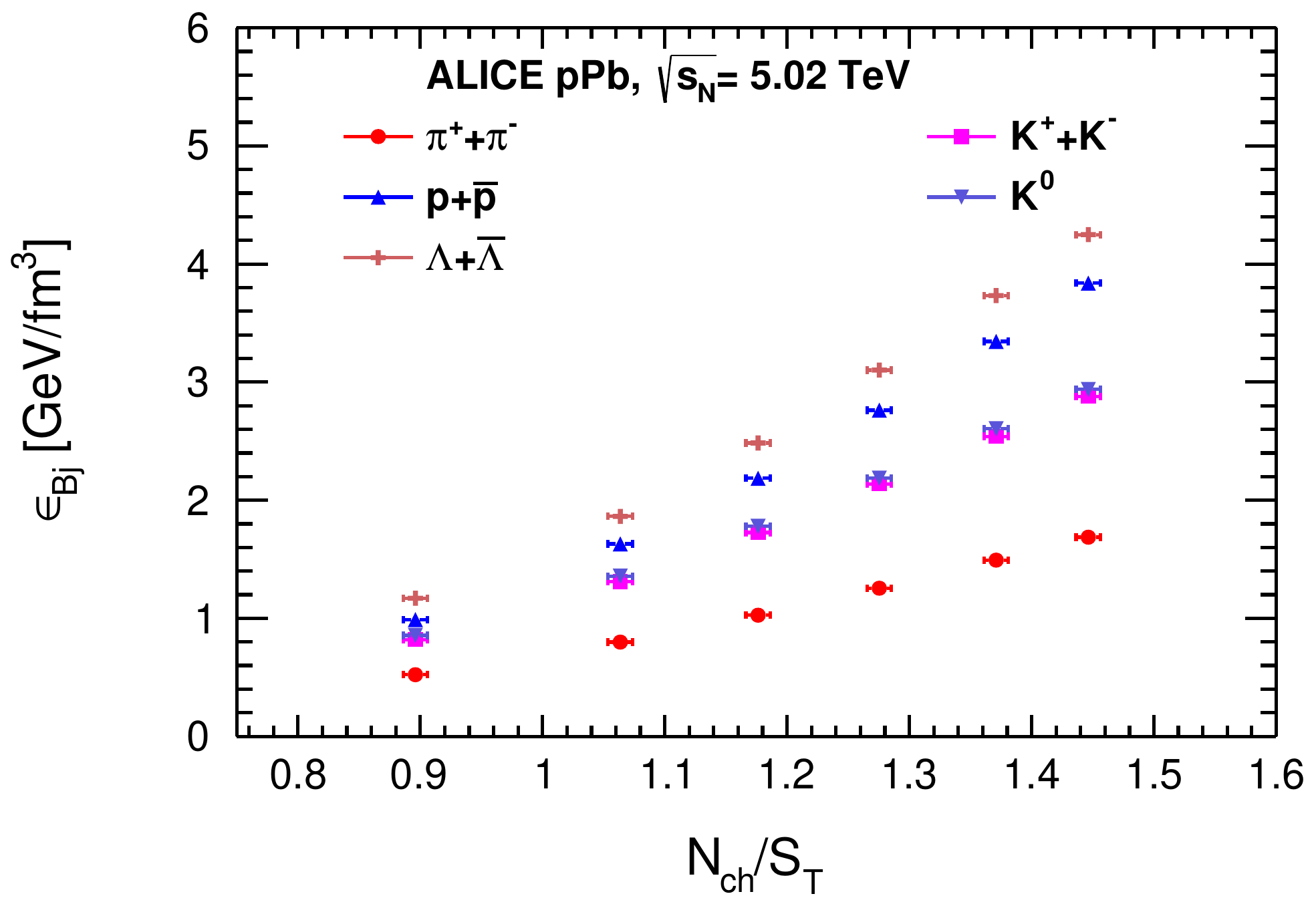}
		\caption{} \label{fig:7b}
	\end{subfigure}% 
	\caption{Bjorken energy density for charged hadrons (a) from different colliding systems and energies as a function of multiplicity scaled by the overlap transverse collision area for \alice data ~\cite{Abelev:2013bla}. The same quantity for identified particles from \alice data~\cite{ALICE:2013wgn} is shown in panel (b).}
	\label{fig:7}       % Give a unique label
\end{figure}

\noindent
The  entropy density ($\sigma$) is determined  in statistical QCD with  dynamical quarks~\cite{Redlich:1985uw} as:

\begin{equation}
\label{Eq.EntropyDensity}
\sigma \approx  \epsilon_{Bj}/\apt,
\end{equation}
\noindent
where the initial Bjorken energy density is given by  Eq.~(\ref{Eq.Bjorken}).
\noindent
This means that the entropy density only depends on the initial energy density and therefore on the multiplicity reached in the collisions. The previous equations can also be written as\\

\begin{eqnarray}
\label{EnergyEntropy}
\frac{\epsilon_{Bj}}{ \langle p_T \rangle ^{4} }  \equiv  \frac{\sigma}{  \langle p_T \rangle ^{3}}.
\end{eqnarray}

\noindent
Using the average transverse momentum distributions as a function of the multiplicity from \alice and \cms,  we get the energy density of Eq.~(\ref{Eq.Bjorken}).  To investigate possible phase transition in the hadronic matter created in  \pp, \ppb, and \pbpb collisions, we use Eq.~(\ref{EnergyEntropy}).
As the first approximation, using the Color String Percolation Model, the  \apt is proportional to the initial temperature $T$~\cite{Scharenberg:2010zz}. This quantity has also been extracted from an analysis of the average transverse momentum, using a linear relationship, $\apt = a+b T$, between \apt and $T$\cite{Waqas:2019bnc}, where  $a$ and $b$  are free fit parameters. It is worth mentioning that the relation between \apt and $T$  depends on the model used, and 
$b$ is a function of the collision energy. However, as a first approach, we consider \apt  $\sim T$, consequently the Eq.~(\ref{EnergyEntropy}) as a function of \apt provides results similar to those obtained from a plot of $\sigma/T^3$ as a function of $T$, analyzed using lattice QCD~\cite{HotQCD:2014kol}.

\begin{figure}[h!]
	\begin{subfigure}{0.49\textwidth}
		\includegraphics[width=\linewidth]{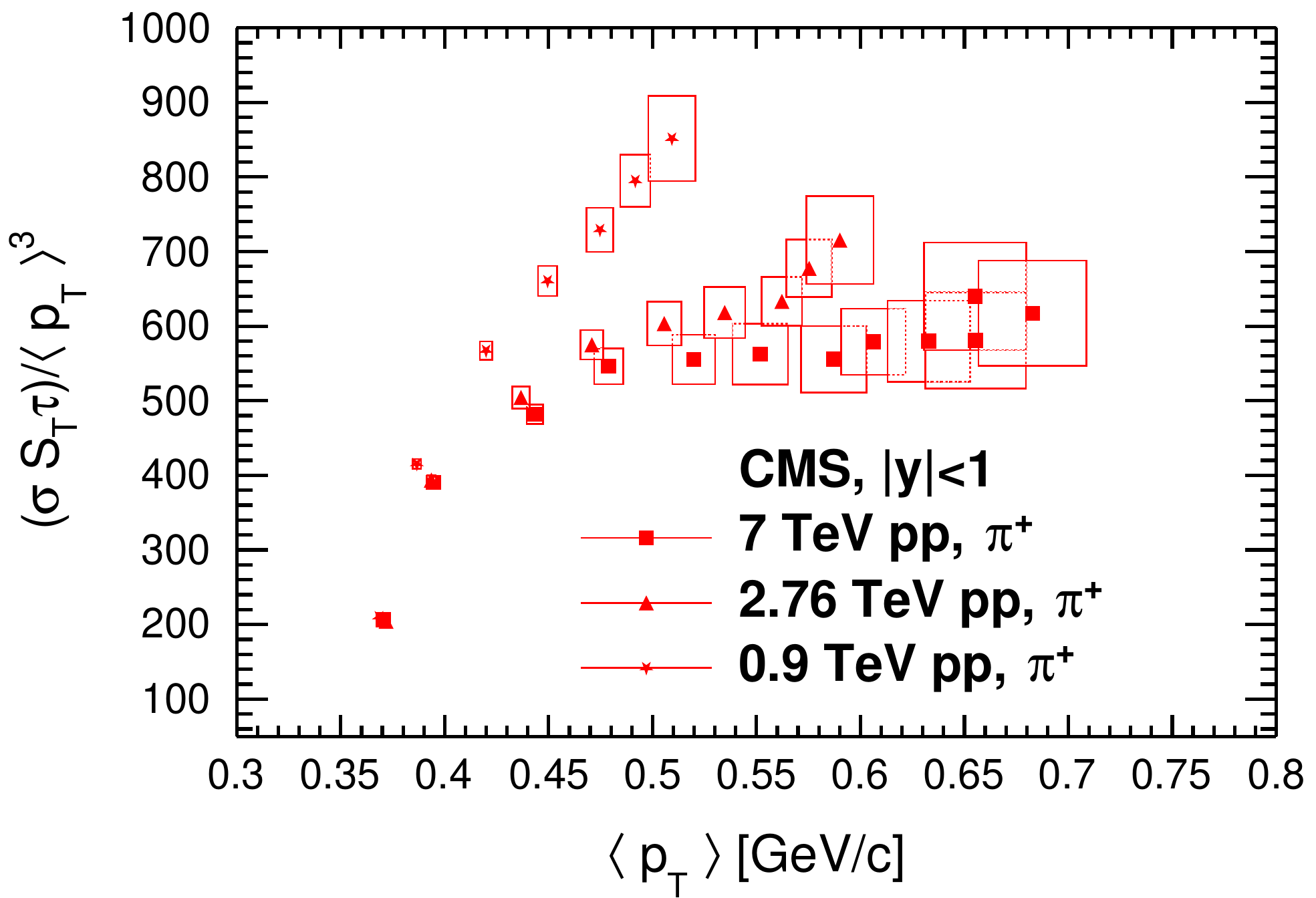}
		\caption{} \label{fig:8a}
	\end{subfigure}%
	\hspace*{\fill}   % maximize separation between the subfigures
	\begin{subfigure}{0.49\textwidth}
		\includegraphics[width=\linewidth]{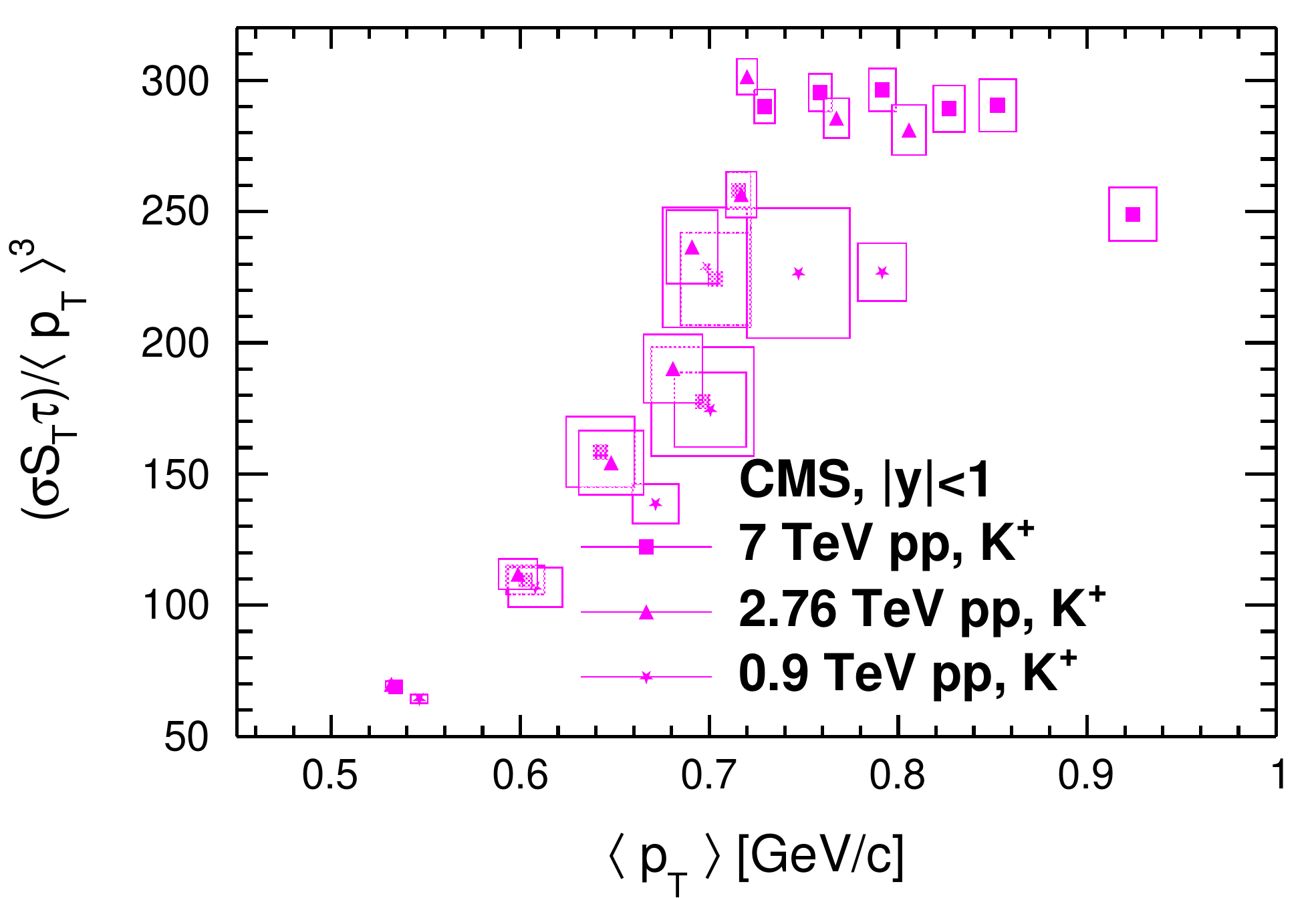}
		\caption{} \label{fig:8b}
	\end{subfigure}% 
	\caption{Entropy density normalized to $\apt^3$ for piones (left) and Kaons (right) from \cms \pp data\cite{CMS:2012xvn} at different energies}
	%    , compared to \ppb data at 5.02 \tev.}
	\label{fig:8}
\end{figure}

\begin{figure}[h!]
	\begin{subfigure}{0.49\textwidth}
		\includegraphics[width=\linewidth]{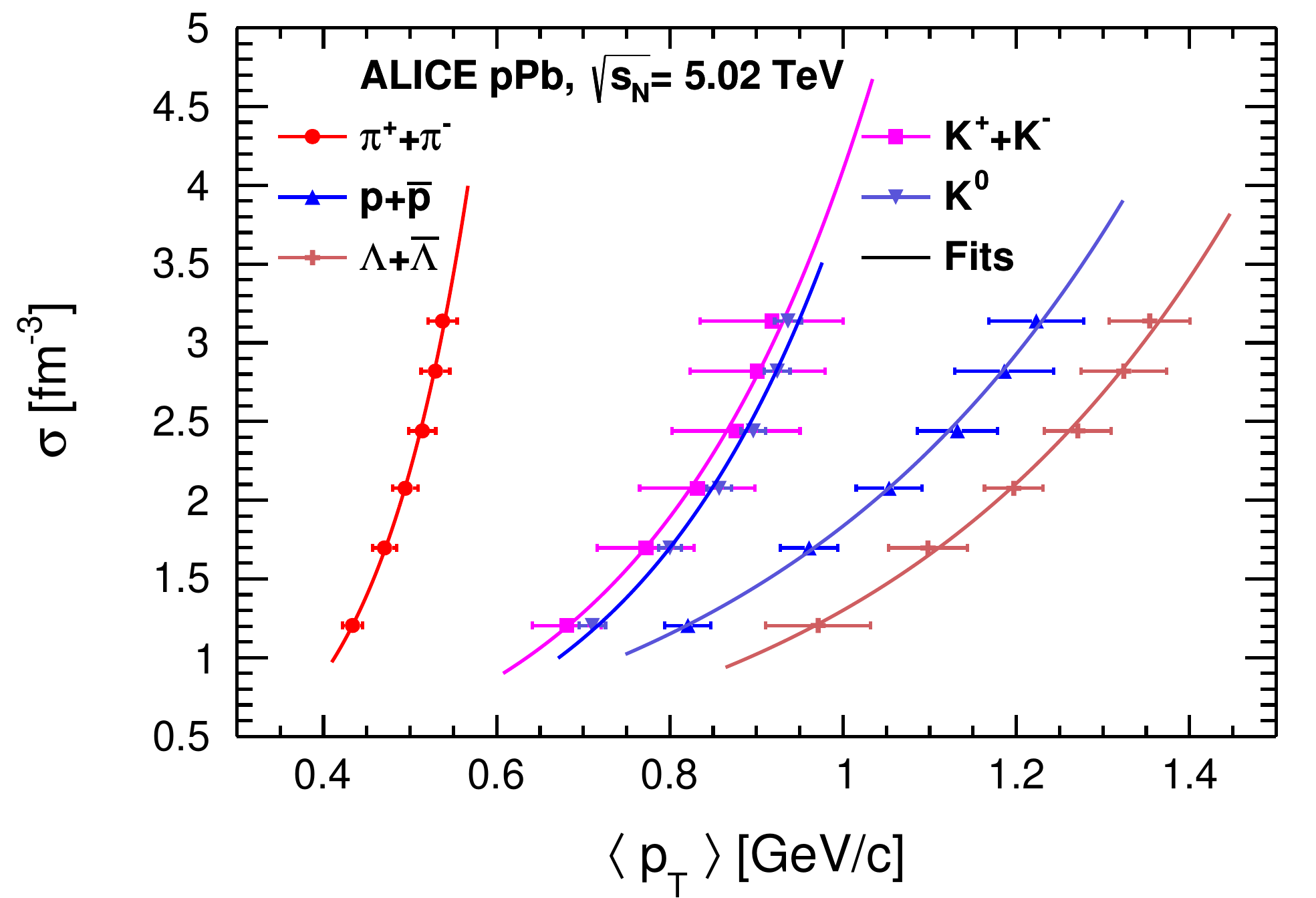}
		\caption{} \label{fig:9a}
	\end{subfigure}%
	\hspace*{\fill}   % maximize separation between the subfigures
	\begin{subfigure}{0.49\textwidth}
		\includegraphics[width=\linewidth]{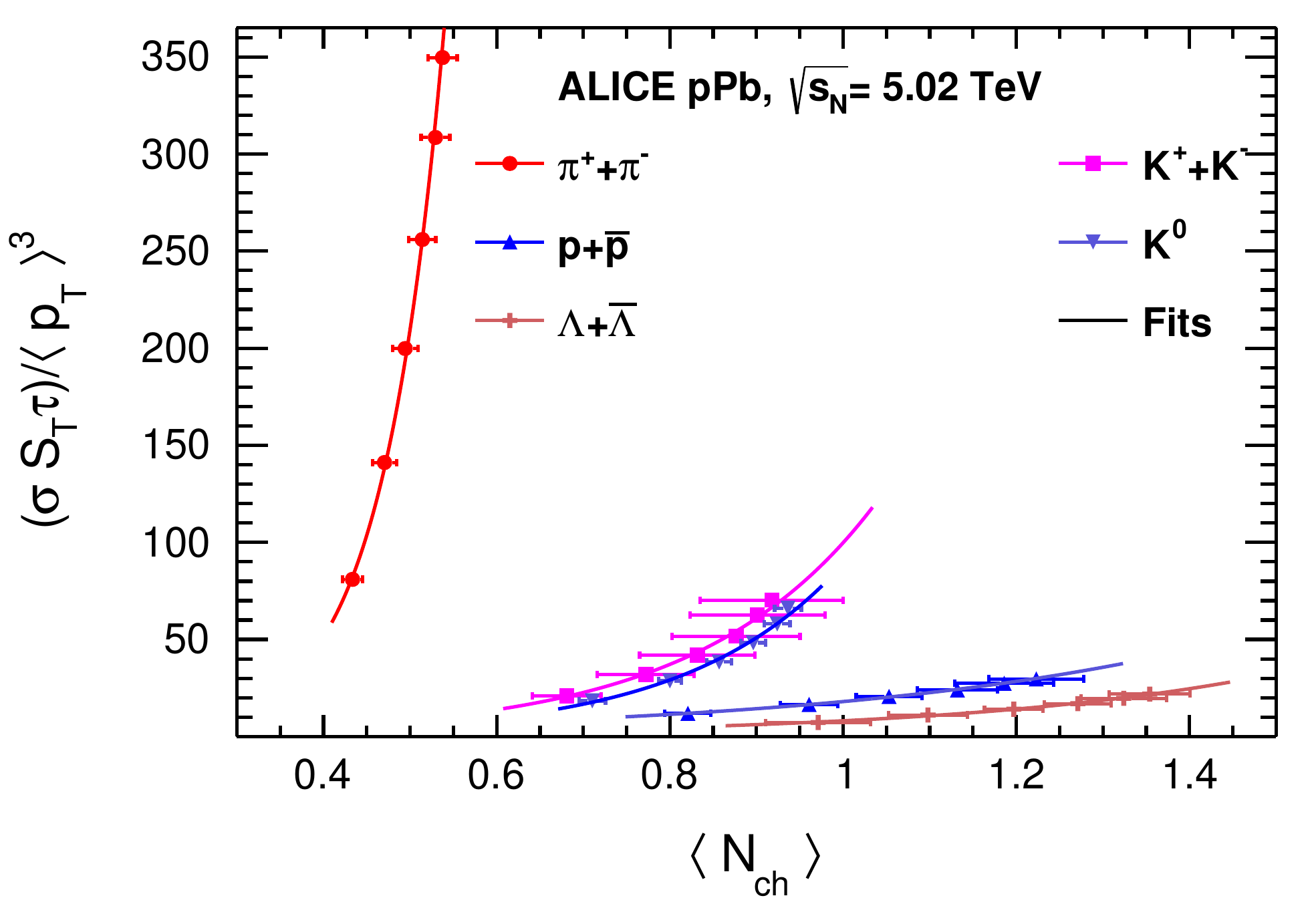}
		\caption{} \label{fig:9b}
	\end{subfigure}% 
	\caption{Entropy density as a function of \apt for identified particles (a). Entropy normalized to $\apt^3$ as a function of \apt, (b). Both cases, correspond to \alice  \ppb data~\cite{ALICE:2013wgn} at different energies.}
	\label{fig:9}  
\end{figure}

\noindent
Figure ~\ref{fig:8} shows the plot of entropy scaled by $\apt^3$ ($\sigma\cdot S_T\tau/\apt^{3}$)  as a function of \apt computed using \cms data~\cite{CMS:2012xvn}. The distributions correspond to  pions (Fig.~\ref{fig:8} (a)) and kaons ( Fig.~\ref{fig:8} (b)) from  \cms \pp collisions at 0.9, 2.76, 7.0 \tev \cite{CMS:2012xvn}. The plot of the entropy instead of entropy density is to avoid the use of large uncertainty radii extracted from \cms, which precludes possible energy effects. This entropy has a rapid increase and seems to be saturated  for pions and kaons produced in \pp at 7 \tev.
Significant fluctuations in the behavior of entropy for the same species at different energies could be associated with the selection processes in the measurements by the \cms experiment.
\noindent
Figure ~\ref{fig:9} (a) shows our results for entropy density as a function of \apt for pions, kaons, protons, and Lambda's,  computed with  \ppb \alice  data~\cite{ALICE:2013wgn}. In all cases, the entropy rises to reach the values determined by the multiplicity as a function of \apt; we observe a faster growth for the lightest particles. When the entropy (entropy density multiplied by $S_{T}\tau$) is normalized to $\apt^3$ (Fig.~\ref{fig:9} (b)), however,  all the cases can be fitted to a functional form  $Exp$(A+B\apt) where A and B  are shown in the Table~\ref{Tparameters}.  The distributions for $\Lambda$'s and protons can also be fitted with a linear functional form, but the best fits are for exponential form.

%\small\begin{verbatim}
\begin{table}[h!]
\caption{\label{Tparameters} Parameters of the fitted  function: $ Exp$ (A +B \apt), in   Fig.~\ref{fig:9} (a) and ~\ref{fig:9} (b).}
\begin{indented}
\lineup
 \item[]\begin{tabular}{@{}*7{l}}
\br
     & &$\pi^{+} +\pi^{-}$ &   $K^{+} +K^{-}$ &  $p^{+} +\bar{p}$  & $K^{0} $ & $\Lambda +\bar{\Lambda}$ \cr
\mr
Fig. 9 (a) & A & -3.7114 & -2.4520 & -1.7269 & -2.7737 & -2.1473\cr
        & B [1/(\gevc)] & 8.9957  & 3.8628  & 2.3344  & 4.1288 & 2.4104 \cr
\mr
Fig. 9 (b) & A & -1.6948 & -0.3344 & 0.6161 & -1.0888 & -0.6768\cr
        & B [1/(\gevc)]  & 14.0796 & 4.9379   & 2.2758  &  5.5768 &  2.7725\cr
\br
\end{tabular}
\end{indented}
\end{table}
%\end{verbatim}\normalsize

\begin{figure}[h!]
		\begin{subfigure}{0.33\textwidth}
		\includegraphics[width=\linewidth]{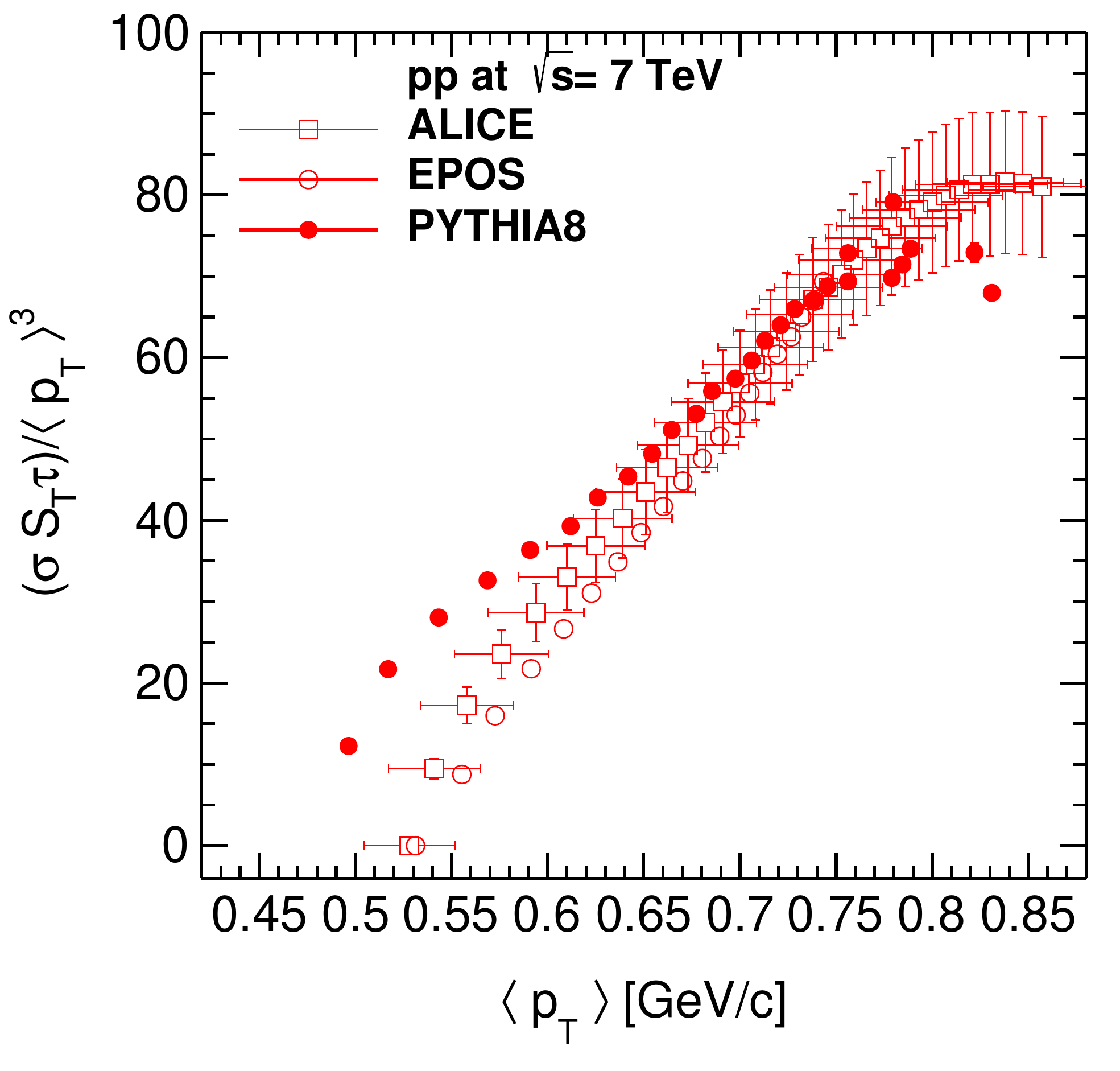}
		\caption{} \label{fig:10a}
	\end{subfigure}%
	\hspace*{\fill}   % maximize separation between the subfigures
	\begin{subfigure}{0.33\textwidth}
		\includegraphics[width=\linewidth]{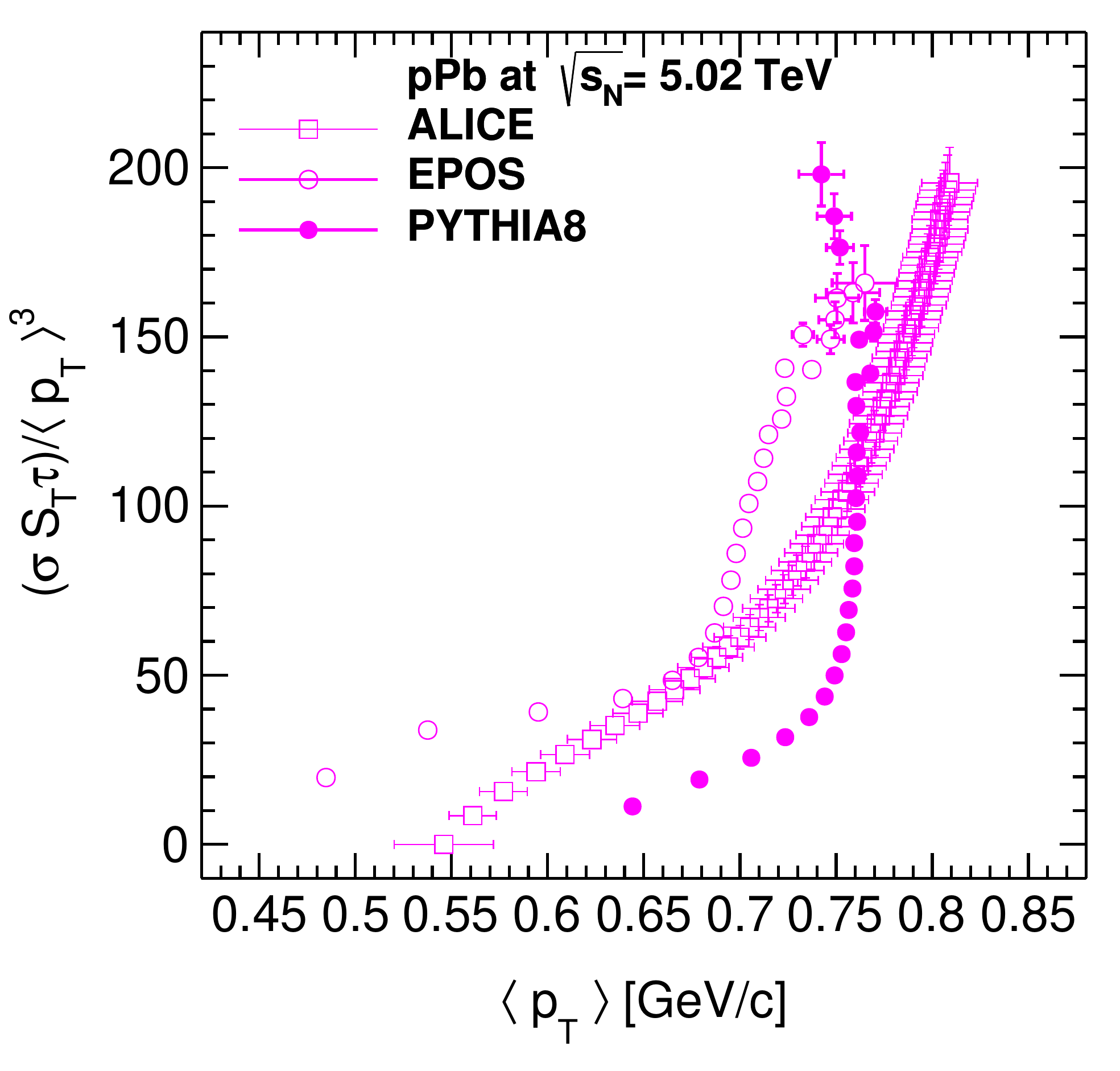}
		\caption{} \label{fig:10b}
	\end{subfigure}%
	\hspace*{\fill}   % maximizeseparation between the subfigures
	\begin{subfigure}{0.33\textwidth}
		\includegraphics[width=\linewidth]{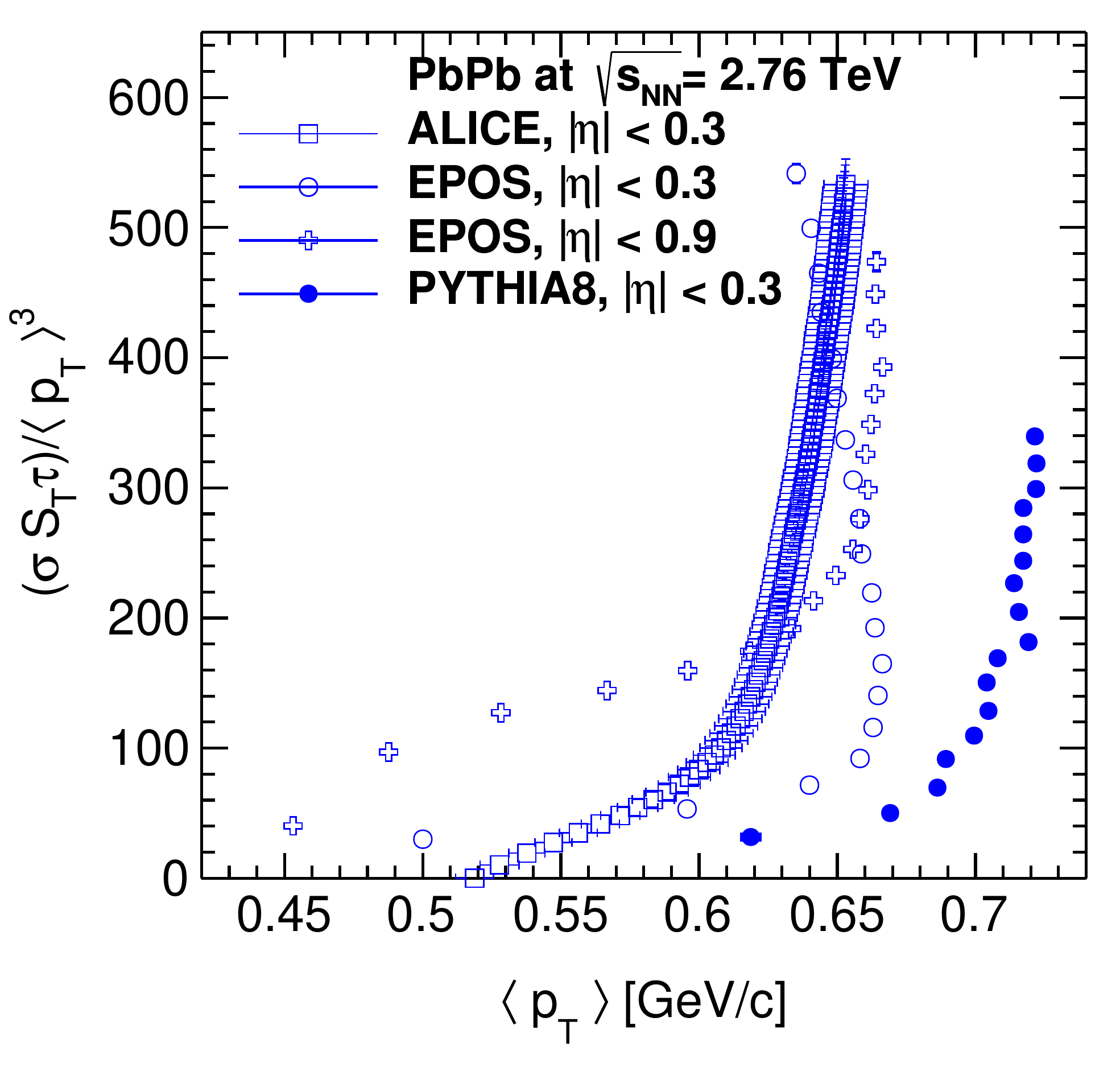}
		\caption{} \label{fig:10c}
	\end{subfigure}
	\caption{Entropy density normalized to $\apt^3$ for \alice \pp, \ppb and  \pbpb data  ~\cite{Abelev:2013bla} at different energies}
	\label{fig:10}%.c6  
\end{figure}

\noindent
Finally, the  results for the entropy scaled to $\apt^3$ for all charged particles produced in \alice \pp, \ppb, and \pbpb colliding systems are shown in Fig.~\ref{fig:10}, together with predictions from the Monte Carlo event generators for the same systems and energies as the \alice data. Figure ~\ref{fig:10} (a) shows the experimental data from \pp collisions, which are well described by \epos, and  \pythia predictions. It is interesting to note a  kind of saturation for \apt larger than 0.8 \gevc.\\ Regarding \ppb collisions,  Fig. ~\ref{fig:10} (b) shows that data and simulation have the same trend:  a rapid growth of the slope that appears  around $\apt \approx 0.7$ \gevc and $\apt \approx 0.75$ \gevc for \epos and \pythia respectively,  while data are between both event generators. We emphasize  that a well defined inflection point, around \apt = 0.62 \gevc,  is observed for \epos when simulations cross the experimental data.\\
For \pbpb colliding system, shown in Fig. 10 (c), a sudden change in the entropy is observed in data and almost the same trend for \pythia is observed but shifted to higher \apt values. The \epos event generator produces a larger slope such that the distribution crosses the data and seems to bend backwards because the \apt rises and then falls as the multiplicity increases, in the pseudorapidity range of $\vert \eta \vert < 0.3$. This value was used to compare with the \alice data. Figure 10 (c), also shows the same distribution produced with \epos in $\vert \eta \vert < 0.9$. We  notice that it rises with a slope closer to that of  data. The results observed from \epos suggest that caution has to be taken to describe the most central pseudorapidity region, since this event generator does not describe data. 
%\noindent
Looking at the details of the sudden change in the entropy $ \equiv \sigma  \ST \tau $ normalized to $\apt^3$ as a function of \apt for the \pbpb \alice data, we observe almost a linear growth for  \apt less than $\approx 0.61$ \gevc, while an exponential growth is observed for   \apt larger than 0.61 \gevc. Assuming $\apt = 3T$~\cite{Gardim:2019xjs}, the entropy normalized to temperature ($S/T^3$) has a linear behavior with respect to the temperature in the range from 0.172 to 0.203 \gevc, and an exponential one for temperatures larger than 0.203 \gevc.\\
Within our  approaches we have obtained  properties of the medium created in different colliding systems. More details can be computed; however, any conclusion from this analysis can only have a limited agreement with data and inspired  QCD models due to the simplified model we used. Other properties will be reported elsewhere.

%%%%%%%%%%%%%%%%%%%%%%%%%%%%%%%%%%%%%%%%%%%%%%%%%%%%%%%%%%%%%
\section{Conclusions and remarks}\label{conclusions}

The first part of this work presents an analysis of the correlation between transverse momentum and multiplicity distributions for different colliding systems at energies of 0.01, 0.9, 2.76, 5.0, 7.0, and 13 \tev using two event generators: \pythia with/without  CR;  \epos with/without fusion model as part of the hadronization. The second part of the work carried out a data analysis of the average transverse momentum versus multiplicity from \alice and \cms experiments to extract thermodynamical properties of the system created in \pp, \ppb, and \pbpb collisions at different collision energies. The results were compared with predictions from \epos and \pythia event generators.\\
Average multiplicity distributions increase with the collision energy  and can be parameterized as  $\aNch = a +b \cdot \ptt^c$, where $a, b$, and $c$, are free parameters for each energy. The \aNch is observed to rise more rapidly with increasing the transverse momentum for higher collision energies.
The hadronization mechanism incorporated through  CR in \pythia produces a more substantial reduction in multiplicity than the \epos fusion model; this effect is more visible for transverse momentum larger than 0.5 \gevc. On the other hand, the average transverse momentum as a function of the multiplicity, computed with \pythia, shows that a differential analysis in terms of impact parameter is not enough to explain the data. 
The CR with different strengths is enough to explain the behavior of data at collision energies of 0.9, 2.76, and 7 \tev reported by the \alice experiment. This  has been tested for other energies finding good agreement. Nonetheless, in the case of the \epos fusion model with  a combination of impact parameters, which is related to  hadronic matter model in the proton,  it allows an excellent description of the data, although \epos results without cut on impact parameter are within  experimental uncertainties.\\
We have shown that the average transverse momentum and Bjorken energy density as a function of multiplicity, normalized to the transverse interaction area, present a scaling law. The results were computed with \alice data, \pythia, and \epos event generators. For all cases,
the \apt scaling breaks in the lowest and highest multiplicity values, which can not be associated with flow effects, as other authors have suggested.\\
Related to thermodynamical properties obtained with  an analysis of  \cms and \alice data: the results indicate an exponential growth of the energy and entropy normalized to $\apt^3$ as a function of the average transverse momentum for charged and identified particles measured in \pp, \ppb, and \pbpb collisions at the available energies. This growth is faster for light particles with respect to heavy ones.
Considering \apt as a linear function of the initial temperature of the system created in the collisions, the conclusions obtained from $\epsilon_{Bj}/\apt^4$ are similar to those from  $\epsilon_{Bj}/T^4$, a quantity widely studied to look for a QCD phase transition. The same conclusion can be reached from  $\sigma/T^3$.
Analysis of experimental \pbpb data with  $\apt = 3T$ produces an effective temperature range of ~ 0.17-0.2 \gevc, where the energy density normalized to $T^4$ has a linear behavior, and for effective temperatures larger than 0.2 \gevc, it grows exponentially. Our results from \epos and \pythia \pp collisions agree with \alice \pp data and have the behavior obtained by lattice (2+1) flavor QCD. They show a kind of saturation for \apt larger than 0.8 \gevc, which is not observed for either \ppb and \pbpb \alice data. 
The predictions from \pythia and \epos show the data trend, but there is no complete agreement to \alice data. Particularly, the \apt in \pbpb from \epos does not describe  the data  at the most central pseudorapidity and consequently produces two values of the entropy density for one \apt value, which is not a physical result.\\
Results for other thermodynamical variables require a more detailed analysis of the relation between \apt and temperature and more precision on the radii measurements to look for a quantitative possible QCD phase transition.
It is important to remember that, the results on the energy and entropy densities are calculated using simple models of Landau and Bjorken, respectively,  based on hydrodynamical approaches. Consequently, our conclusions can thus be considered as qualitative results that could be improved using more realistic models in the calculations, hadronic matter in the proton, and models for the core created in \pp collisions.\\

%%%%%%%%%%%%%%%%%%%%%%%%%%%%%%%%%%%%%%%%%%%%%%%%%%%%%%%%%%%%%
\section*{Acknowledgements}
Partial support was received by DGAPA-PAPIIT IG100219, IG100322 and CONACyT A1-S-16215 and A1-S-13525 projects. M.R.C. thankfully acknowledge computer resources, technical advise and support provided by Laboratorio Nacional de Super c\'omputo del Sureste de M\'exico (LNS), a member of the CONACYT national laboratories, with project No. 53/2017. The authors  thank to L. D\'iaz  by the technical support, and A. Ayala for  helpful discussions and comments.
\section*{References}
\bibliographystyle{iopart-num.bst}

\bibliography{Manuscript_JPGV9_rev}

\end{document}